\documentstyle[12pt,epsf]{article} \hoffset -0.5in \textwidth 6.00in
\textheight
8.5in  \parskip 7pt \openup1.5\jot \parindent=0.5in
\newskip\humongous \humongous=0pt plus 1000pt minus 1000pt
\def\caja{\mathsurround=0pt}
\def\eqalign#1{\,\vcenter{\openup1\jot \caja
        \ialign{\strut \hfil$\displaystyle{##}$&$
        \displaystyle{{}##}$\hfil\crcr#1\crcr}}\,}
\newif\ifdtup

\def\eqright #1\cr{\noalign{\hfill$\displaystyle{{}#1}$}}
\def\eqleft #1\cr{\noalign{\noindent$\displaystyle{{}#1}$\hfill}}

\def\oldreffmt#1{\rlap{[#1]} \hbox to 2\parindent{}}

\def\figfmt#1{\rlap{Figure {#1}} \hbox to 1in{}}

\def\VEV#1{\left\langle #1\right\rangle}

\def\sectioneq{\def\theequation{\thesection.\arabic{equation}}{\let
\holdsection=\section\def\section{\setcounter{equation}{0}\holdsection}}}%

\def\auto{\eqno(\refstepcounter{equation}\theequation)}
\def\begineq #1\endeq{$$ \refstepcounter{equation}\eqalign{#1}\eqno
	(\theequation) $$}
\def\contlimit{\,{\hbox{$\longrightarrow$}\kern-1.8em\lower1ex
\hbox{${\scriptstyle (a\rightarrow0)}$}}\,}
\def\centeron#1#2{{\setbox0=\hbox{#1}\setbox1=\hbox{#2}\ifdim
\wd1>\wd0\kern.5\wd1\kern-.5\wd0\fi
\copy0\kern-.5\wd0\kern-.5\wd1\copy1\ifdim\wd0>\wd1
\kern.5\wd0\kern-.5\wd1\fi}}
\def\centerover#1#2{\centeron{#1}{\setbox0=\hbox{#1}\setbox
1=\hbox{#2}\raise\ht0\hbox{\raise\dp1\hbox{\copy1}}}}
\def\centerunder#1#2{\centeron{#1}{\setbox0=\hbox{#1}\setbox
1=\hbox{#2}\lower\dp0\hbox{\lower\ht1\hbox{\copy1}}}}
\def\lsim{\;\centeron{\raise.35ex\hbox{$<$}}{\lower.65ex\hbox
{$\sim$}}\;}
\def\gsim{\;\centeron{\raise.35ex\hbox{$>$}}{\lower.65ex\hbox
{$\sim$}}\;}
\def\st#1{\centeron{$#1$}{$/$}}

\def\super#1{\ifmmode \hbox{\textsuper{#1}}\else\textsuper{#1}\fi}
\def\textsuper#1{\newcount\holdspacefactor\holdspacefactor=\spacefactor
$^{#1}$\spacefactor=\holdspacefactor}

\def\getcite#1,{\advance\citenumber by1
\def\getcitearg{#1}\def\lastarg{@}
\ifnum\citenumber=1
\ref{#1}\let\next=\getcite\else\ifx\getcitearg\lastarg\let\next=\relax
\else ,\ref{#1}\let\next=\getcite\fi\fi\next}

\def\pom{{\rm P\kern -0.53em\llap I\,}}
\def\spom{{\rm P\kern -0.36em\llap \small I\,}}
\def\sspom{{\rm P\kern -0.33em\llap \footnotesize I\,}}

\relax

\newskip\humongous \humongous=0pt plus 1000pt minus 1000pt
\def\caja{\mathsurround=0pt}
\def\eqalign#1{\,\vcenter{\openup1\jot \caja
        \ialign{\strut \hfil$\displaystyle{##}$&$
        \displaystyle{{}##}$\hfil\crcr#1\crcr}}\,}
\newif\ifdtup

\def\eqright #1\cr{\noalign{\hfill$\displaystyle{{}#1}$}}
\def\eqleft #1\cr{\noalign{\noindent$\displaystyle{{}#1}$\hfill}}

\def\oldreffmt#1{\rlap{[#1]} \hbox to 2\parindent{}}

\def\figfmt#1{\rlap{Figure {#1}} \hbox to 1in{}}

\def\VEV#1{\left\langle #1\right\rangle}

\def\auto{\eqno(\refstepcounter{equation}\theequation)}
\def\begineq #1\endeq{$$ \refstepcounter{equation}\eqalign{#1}\eqno
	(\theequation) $$}
\def\contlimit{\,{\hbox{$\longrightarrow$}\kern-1.8em\lower1ex
\hbox{${\scriptstyle (a\rightarrow0)}$}}\,}
\def\centeron#1#2{{\setbox0=\hbox{#1}\setbox1=\hbox{#2}\ifdim
\wd1>\wd0\kern.5\wd1\kern-.5\wd0\fi
\copy0\kern-.5\wd0\kern-.5\wd1\copy1\ifdim\wd0>\wd1
\kern.5\wd0\kern-.5\wd1\fi}}
\def\centerover#1#2{\centeron{#1}{\setbox0=\hbox{#1}\setbox
1=\hbox{#2}\raise\ht0\hbox{\raise\dp1\hbox{\copy1}}}}
\def\centerunder#1#2{\centeron{#1}{\setbox0=\hbox{#1}\setbox
1=\hbox{#2}\lower\dp0\hbox{\lower\ht1\hbox{\copy1}}}}
\def\lsim{\;\centeron{\raise.35ex\hbox{$<$}}{\lower.65ex\hbox
{$\sim$}}\;}
\def\gsim{\;\centeron{\raise.35ex\hbox{$>$}}{\lower.65ex\hbox
{$\sim$}}\;}
\def\st#1{\centeron{$#1$}{$/$}}

\def\super#1{\ifmmode \hbox{\textsuper{#1}}\else\textsuper{#1}\fi}
\def\textsuper#1{\newcount\holdspacefactor\holdspacefactor=\spacefactor
$^{#1}$\spacefactor=\holdspacefactor}

\def\getcite#1,{\advance\citenumber by1
\ifnum\citenumber=1
\ref{#1}\let\next=\getcite\else\ifx#1@\let\next=\relax
\else ,\ref{#1}\let\next=\getcite\fi\fi\next}

\def\upon #1/#2 {{\textstyle{#1\over #2}}}
\relax

\def\mainhead#1{\setcounter{equation}{0}\addtocounter{section}{1}
  \vbox{\begin{center}\large\bf #1\end{center}}\nobreak\par}
\sectioneq
\def\subhead#1{\bigskip\vbox{\noindent\bf #1}\nobreak\par}

\def\til#1{\centeron{\hbox{$#1$}}{\lower 2ex\hbox{$\char'176$}}}
\def\tild#1{\centeron{\hbox{$\,#1$}}{\lower 2.5ex\hbox{$\char'176$}}}
\def\sumtil{\centeron{\hbox{$\displaystyle\sum$}}{\lower
-1.5ex\hbox{$\widetilde{\phantom{xx}}$}}}

\def\pom{{\rm P\kern -0.53em\llap I\,}}
\def\spom{{\rm P\kern -0.36em\llap \small I\,}}
\def\sspom{{\rm P\kern -0.33em\llap \footnotesize I\,}}

\newcommand{\bit}{\begin{itemize}}
\newcommand{\eit}{\end{itemize}}

\newcommand{\beq}{\begin{equation}}
\newcommand{\eeq}{\end{equation}}
\newcommand{\beqa}{\begin{eqnarray}}
\newcommand{\eeqa}{\end{eqnarray}}

\begin{document}

\begin{titlepage}

\rightline{\vbox{\halign{&#\hfil\cr
&ANL-HEP-PR-95-83\cr
&UF-IFT-HEP-95-22\cr}}}

\vspace{.4in}

\begin{center}

{\bf $t$-CHANNEL UNITARITY CONSTRUCTION

OF SMALL-x KERNELS}\footnote{Lectures
presented by Alan R. White at XXXVth Cracow School of
Theoretical Physics ``Bozonization and Conformal Symmetry in High Energy and
Condensed Matter Physics'', Zakopane, Poland (June 1995).}\footnote{Work
supported by the U.S. Department of Energy, Division of High
Energy Physics, \newline Contracts W-31-109-ENG-38 and DEFG05-86-ER-40272}

\medskip

{Claudio Corian\`{o}$^{a,b}$
\footnote{coriano@phys.ufl.edu ~$^{\#}$arw@hep.anl.gov}
and \ Alan. R. White$^{a\#}$}

\vskip 0.6cm

\centerline{$^a$High Energy Physics Division}
\centerline{Argonne National Laboratory}
\centerline{9700 South Cass, Il 60439, USA.}
\vspace{0.5cm}

\centerline{$^b$Institute for Fundamental Theory}
\centerline{Department of Physics}
\centerline{ University of Florida at Gainesville, FL 32611, USA}
\vspace{0.5cm}

\end{center}

\begin{abstract}

We present the BFKL equation as a reggeon Bethe-Salpeter equation and
discuss the use of reggeon diagrams to obtain 2-2 and 2-4 reggeon interactions
at $O(g^4)$. We then outline the dispersion theory basis of multiparticle
$j$-plane analysis and describe how a gauge theory can be studied by
combining Ward identity constraints with the group structure of reggeon
interactions. The derivation of gluon reggeization, the $O(g^2)$ BFKL
kernel, and $O(g^4)$ corrections, is described within this formalism.
We give an explicit expression for the $O(g^4)$ forward ``parton'' kernel in
terms of logarithms and evaluate the eigenvalues. A separately infra-red
finite component with a holomorphically factorizable spectrum is shown to be
present and conjectured to be a new leading-order partial-wave amplitude. A
comparison is made with Kirschner's discussion of $O(g^4)$ contributions
from the multi-Regge effective action.

\end{abstract}

\end{titlepage}

\mainhead{1. INTRODUCTION}

In the leading-log approximation, the small-x behavior of parton
distributions in QCD is derived from the BFKL evolution equation\cite{bfkl}.
It is well-known that the BFKL kernel is (and was derived as) a 2-2 reggeon
interaction - with the reggeon being a reggeized gluon. For general $t~(=
-q^2)$ the BFKL equation becomes a reggeon
Bethe-Salpeter equation as illustrated in Fig.~1.1
\begin{center}
\leavevmode
\epsfxsize=4in
\epsffile{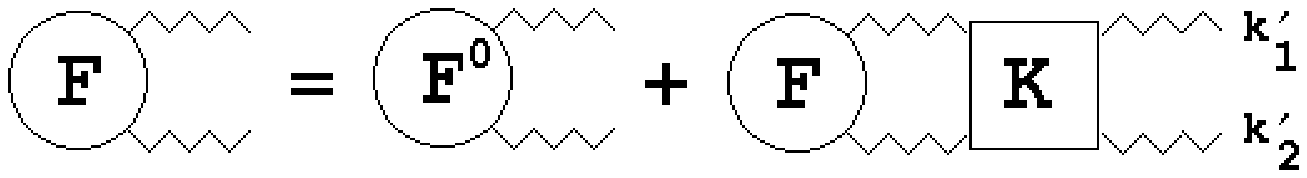}

Fig.~1.1 Reggeon Bethe-Salpeter equation
\end{center}
where $F~\equiv~F(\omega, k_1,k_2)$ is (the fourier transform of) a
two-reggeon amplitude which becomes a parton distribution when $k_1+k_2 =
0$. The two-reggeon intermediate state integration is

\noindent\parbox{1.5in}{
\begin{center}
\leavevmode
\epsfxsize=1in
\epsffile{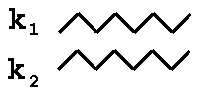}
\end{center}}
\parbox{4.5in}{
$$
\equiv ~~~
\int {d^2k_1 \over k_1^2} {d^2k_2 \over k_2^2} {\delta^2(k_1'+k_2'-k_1-k_2)
\over \omega - \Delta(k_1^2) -\Delta(k_2^2)}
\auto\label{2rp}
$$}
where $\Gamma_2 = [\omega - \Delta(k_1^2) -\Delta(k_2^2)]^{-1}$ is
a two-reggeon propagator and
$$
\eqalign{ \Delta(q^2)= Ng^2 J_1(q^2)~=~{Ng^2 \over 16{\pi}^3}
\int {d^2k \over k^2(k-q)^2} }
\auto\label{j1}
$$
The 2-2 reggeon interaction $K(k_1,k_2,k'_1,k'_1)$ is given by

\noindent \parbox{2in}{
\begin{center}
\leavevmode
\epsfxsize=1.5in
\epsffile{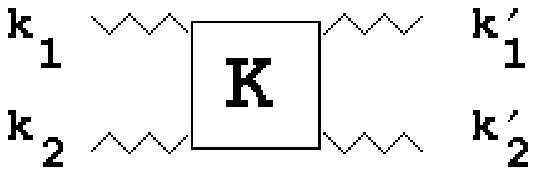}
\end{center}}
\parbox{4in}{
$$
\eqalign{~~=~Ng^2\Biggl( {k_1^2{k_2'}^2~+~k_2^2{k_1'}^2 \over (k_1-{k_1'})^2}
-(k_1+k_2)^2 }\Biggr)
\auto\label{int}
$$}
The familiar BFKL kernel is
$$
\eqalign{K_{BFKL}(k_1,k_2,k'_1,k'_1)~=~& K(k_1,k_2,k'_1,k'_1)\cr
 &-~ \Delta(k_1^2)\delta^2(k_1 -k'_1) ~-~\Delta(k_2^2)\delta^2(k_2 -k'_2) }
\auto\label{bfkl}
$$

Generally reggeon interactions have been studied by $s$-channel unitarity
calculations\cite{bfkl,bar,fl,kls,lip1}. For example, calculation of the BFKL
kernel can be represented schematically as in Fig.~1.2

\begin{center}
\leavevmode
\epsfxsize=5in
\epsffile{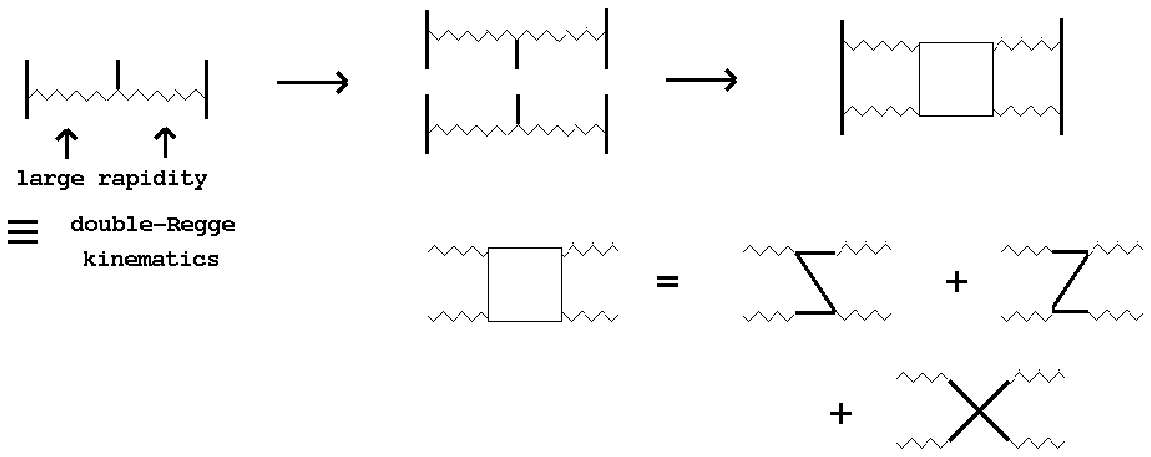}

Fig.~1.2 Calculation of the BFKL Kernel via $s$-channel unitarity.
\end{center}
{\it Alternatively} we can ``sew'' reggeon amplitudes together via
$t$-channel unitarity. The derivation of the BFKL kernel in this manner will
be a core part of these lectures. It is illustrated schematically in
Fig.~1.3.
\begin{center}
\leavevmode
\epsfxsize=5in
\epsffile{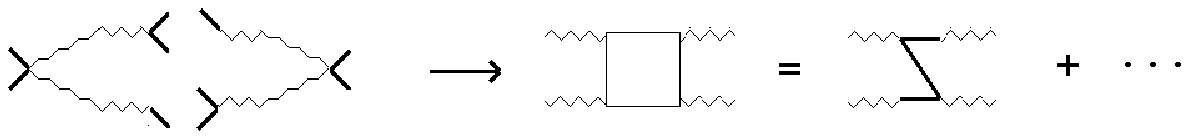}

Fig.~1.3 Calculation of the BFKL Kernel via $t$-channel unitarity.
\end{center}

The ``sewing'' of Fig.~1.3 is well-defined if it is done in the $j$-plane
(where $j=1+ \omega$) by treating the particles appearing in the
intermediate states also as reggeons\cite{cw,uni}. The analytic continuation of
multiparticle unitarity equations in the $j$-plane is a powerful
formalism\cite{gpt,arw1}, essentially because of the underlying exploitation
of multiparticle dispersion theory\cite{harw,hs} involved. We will briefly
describe the full formalism later. First we observe that a simple (but ``too
naive'') way to sew reggeon amplitudes together with reggeons acting as
particles is to use reggeon diagrams directly.

\subhead{1.1 Reggeon Diagrams}

$t$-channel unitarity is satisfied at the level of reggeon unitarity (see
the later discussion) if we construct a set of
reggeon diagrams as follows\cite{ker}. We introduce a triple regge vertex

\noindent \parbox{2in}{
\begin{center}
\leavevmode
\epsfxsize=1in
\epsffile{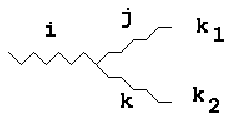}
\end{center}}
\parbox{3in}{
$$
\sim ~~g~c_{ijk}~\sqrt{\alpha'}~~[\omega - \alpha'k_1^2 -
\alpha'k_2^2]
$$}
\parbox{1in}{
$$
{}~\auto\label{trv}
$$}
where $g$ is the gauge coupling, $c_{ijk}$ is a structure
constant color factor, and [...] is a ``nonsense zero''. We introduce
propagators

\noindent \parbox{2in}{
\begin{center}
\leavevmode
\epsfxsize=1.5in
\epsffile{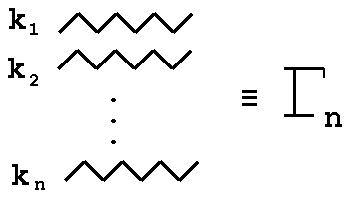}
\end{center}}
\parbox{3in}{
$$
=~~~\Pi_{i=1}^n~\biggl({1 \over \alpha'k_i^2} \biggr)~~ {1 \over \omega
-\sum_{i=1}^n \alpha' k_i^2}
$$}
\parbox{1in}{
$$
{}~\auto\label{prop2}
$$}
We then combine vertices and propagators by integrating over
transverse momenta - with momentum conservation imposed. (A subtlety is that
we actually have to construct ``cut'' reggeon diagrams for the imaginary
part of amplitudes, but we will not elaborate on this).

The nonsense zeroes cancel many reggeon singularities leaving only {\it
particle singularities generating arbitrarily high-order reggeon
interactions}. The outcome is a very simple formalism\cite{ker} for
generating reggeon interactions. The interactions are automatically obtained
in terms of {\it transverse momentum diagrams} which we introduce via the
vertices and phase-space integrations illustrated in Fig.~1.4.

\leavevmode
\epsffile{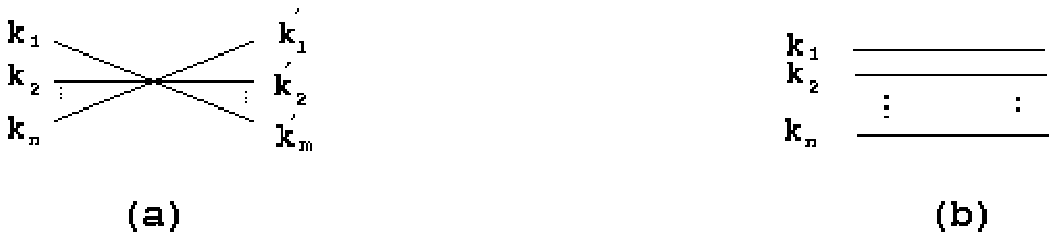}
\begin{center}

Fig.~1.4 (a)Vertices and (b) intermediate states in transverse momentum.
\end{center}
The rules for writing amplitudes corresponding to the diagrams are the
following

\begin{itemize}

\item{For each vertex illustrated in Fig.~1.4(a) we write a factor
$$
16\pi^3\delta^2(\sum k_i~  - \sum k_i')(\sum k_i~)^2
$$
}
\item{For each intermediate state illustrated in Fig.~1.4(b) we write a factor
$$
(16\pi^3)^{-n}\int d^2k_1...d^2k_n~ /~k_1^2...k_n^2
$$}
\end{itemize}
The reggeization, of the gluon, is illustrated in Fig.~1.5

\begin{center}
\leavevmode
\epsfxsize=4in
\epsffile{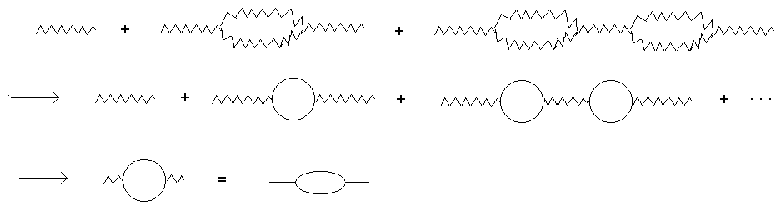}

Fig.~1.5 Reggeization of the gluon from reggeon diagrams
\end{center}
The origin of the BFKL kernel is illustrated in Fig.~1.6

\begin{center}
\leavevmode
\epsfxsize=4in
\epsffile{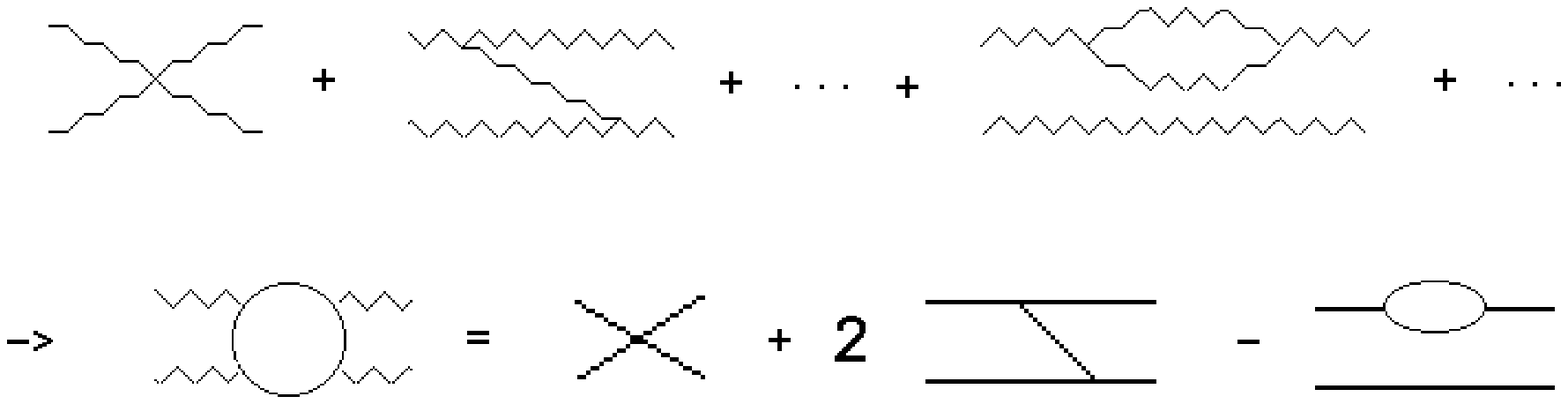}

Fig.~1.6 The BFKL kernel from reggeon diagrams.
\end{center}
and the transverse momentum diagrams derived as $O(g^4)$ interactions
in \cite{ker} are shown in Fig.~1.7.
\begin{center}
\leavevmode
\epsfxsize=4.5in
\epsffile{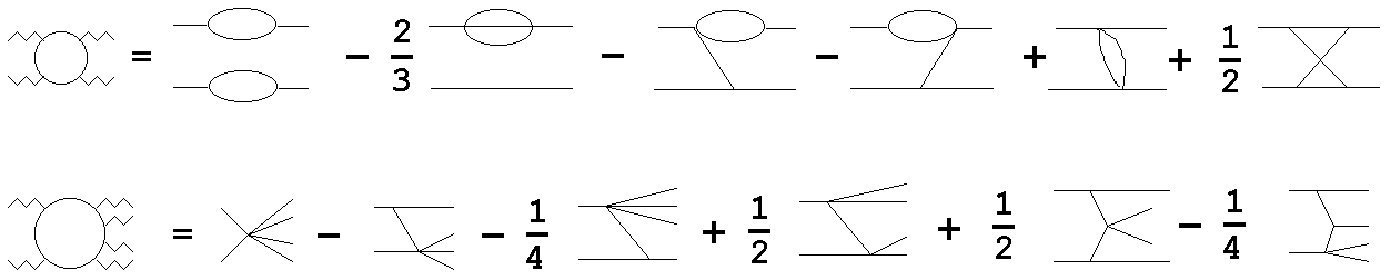}

Fig.~1.7 $O(g^4)$ reggeon interactions.
\end{center}

In all of the above cases it can be shown\cite{ker} that if
reggeon diagrams are used to generate the possible transverse momentum
diagrams then, {\it in color zero channels, gauge invariance determines the
relative coefficients uniquely}. Gauge invariance is imposed by requiring
that

\begin{itemize}

\item{ all infra-red divergences cancel,}

\item{ reggeon interactions vanish when any transverse momentum goes
to zero.}

\end{itemize}
The cancelation of infra-red divergences is, essentially, an obvious
consequence of gauge invariance. As we now elaborate, imposing the vanishing
of reggeon amplitudes at zero tranverse momentum is directly equivalent to
imposing the defining Ward identities of the theory\cite{gth}.

\subhead{1.2 Gauge Invariance and Reggeon Ward Identities}

A reggeon amplitude is defined via a multi-Regge limit in which, say,
$s_i \to \infty$ i=1,..,4. Schematically we can write

\noindent \parbox{2.5in}{\begin{center}
\leavevmode
\epsfxsize=2in
\epsffile{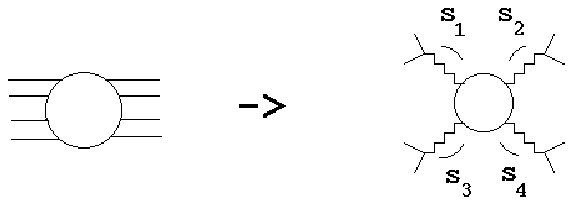}
\end{center}}
\parbox{3.5in}{
$$
\equiv~~\Pi^4_{i=1}
{}~s_i^{\alpha_i}~A_{\alpha_1,\alpha_2,\alpha_3,\alpha_4}
\auto\label{rea2}
$$}
We can always find a Lorentz frame in which the limit $s_1 \to \infty$ is
defined by $p_+ \to \infty, k \to k_{\perp}$ where $p$ and $k$ are as
labelled
in Fig.~1.8.

\begin{center}
\leavevmode
\epsfxsize=4in
\epsffile{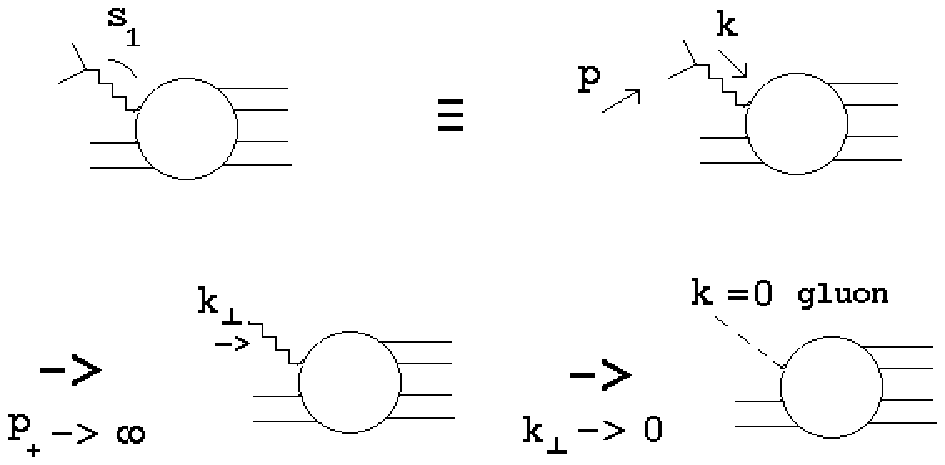}

Fig.~1.8 Reduction of a reggeon amplitude to a gluon amplitude.

\end{center}
The further limit $k_\perp \to 0$ is then equivalent to setting $k=0$.
Reggeization implies the reggeon amplitude must give the $k=0$ gluon amplitude.
Therefore we obtain the zero momentum limit of an amplitude
which satisfies a Ward identity\cite{gth}
$$
k_{\mu}~\VEV{A_{\mu}(k)~...~}~=~0
\auto
$$
where $\VEV{A_{\mu}(k)~...~}$ is the amplitude involving a gluon with momentum
$k_{\mu}$. Differentiating
$$
\eqalign{ &\VEV{A_{\mu}~...~}~+~{\partial \VEV{A_{\nu}~...~}\over
\partial k_{\mu}}~k_{\nu}~=0\cr
=>~&\VEV{A_{\mu}~...~}~~\centerunder{$\to$} {\raisebox{-5mm}
{$(k_{\mu}\to  0)$}}0~~~~~~~if~~ ~{\partial
\VEV{A_{\nu}~...~}\over \partial k_{\mu}}
{}~\st{\to}~\infty}
\auto\label{wd3}
$$
If there are no
internal infra-red divergences occurring explicitly at zero transverse
momentum (as will be the case in the absence of massless
fermions\cite{arw2}), then this identity requires the
amplitude to vanish. Clearly the same argument can be applied to each of the
reggeons in (\ref{rea2}).

\subhead{1.3 Questions}

A number of closely related questions arise from the reggeon diagram
construction of reggeon interaction kernels. We can list some of the more
obvious as follows.

\begin{description}

\item[1.] The kernels are {\it scale-invariant in transverse momentum} -
what is the significance of this ``approximation''?

\item[2.] How is a scale(s) to be added?

a) $g^2/4\pi~~\to ~~\alpha_s(Q^2/\mu^2)$~~?

b) A $k_{\perp}$ cut-off~~?

c) An ``average'' rapidity $\VEV{\eta}$ as a normalization\cite{kir}~~?

\noindent It is generally anticipated that a full next-to-leading order
calculation\cite{fl} will provide an answer to this question.

\item[3.] Why are there only transverse momentum integrals representing
$t$-channel states?

\item[4.] What is the significance of properties related to conformal
invariance?

\end{description}
In the following we will briefly describe a more fundamental derivation of
$O(g^4)$ reggeon interactions directly from $t$-channel unitarity\cite{cw1}.
This formalism provides a solid basis within which to ask these
questions and, at least partly, answer them. A major outcome will be the
suggestion that {\it scale-invariant contributions that are well-defined by
unitarity are necessarily conformally invariant}.

\mainhead{2. MULTIPARTICLE $j$-PLANE ANALYSIS}

To introduce langusge, we first recall the simplest elements of Regge theory
for elastic scattering amplitudes. The partial-wave expansion is
$$
A(z,t)=\sum^\infty_{j=0}(2j+1)a_j(t)P_j(z),\auto\label{3.1}
$$
where
$$
\eqalign{
a_j(t)&={1\over 2}\int^{+1}_{-1}dz A(z,t)P_j(z) }
\auto\label{3.2}
$$
Using the dispersion relation
$$
A(z,t)={1\over2\pi}\int_{I^R + I^L}{dz' \over (z' - z) }\Delta(z',t)
\auto\label{disp}
$$
we obtain
$$
a_j(t)={1\over 4\pi}\int_{I^R + I^L}dz'\Delta(z',t)
\int^{+1}_{-1}{dz \over (z' - z)} P_j(z)
\auto\label{pw}
$$
giving ``signatured'' continuations from even and odd $j$
$$
a^\pm_j(t)=~{1\over2\pi}\int_{I^R}dz'Q_j(z')\Delta(z',t)
\pm (-1)^j{1\over2\pi}\int_{I^L}dz'Q_j(-z')\Delta(z',t).
\auto\label{3.6}
$$
The asymptotic behavior of A(z,t) can be studied via the Sommerfeld-Watson
transform
$$
A(z,t)~=~\sum_{\pm}~\int dj {(2j+1) \over4 sin\pi j}
a^{\pm}_j(t)\biggl(P_j(z)\pm P_j(-z)\biggr)~.
\auto\label{sw}
$$
and a Regge pole in $a^{\pm}_j(t)$ at $j=\alpha(t)$ gives
$$
A(z,t)) ~\sim~z^{\alpha(t)}
\auto\label{rpo}
$$

The simplest example of ``j-plane unitarity'' is elastic unitarity.
$$
a_j^{\tau}~-~a_j^{\tau*}~~=~~i
{}~\rho(t)~a_j^{\tau}a_j^{\tau*}~~~~~~~~~\tau=\pm
\auto\label{elu}
$$
This equation is inconsistent with a fixed pole in the $j$-plane. But
apparently
$$
Q_{j}(z)~ \centerunder{$\longrightarrow$}{\raisebox{-3mm}
{$\scriptstyle j \to -1$}} ~~\Gamma(j +1) ~\sim~ {1 \over j+1}~~~
=>~~~
a^\pm_j(t) ~\centerunder{$\sim$}{\raisebox{-3mm}
{$\scriptstyle j \to -1$}}~~ {1 \over j + 1}
\auto\label{3.8}
$$
and so there is a ``nonsense'' pole at $j~=~n_1~+~n_2~-~1$
where, in this case, $n_1=n_2=0$. In a gauge theory $n_1=n_2=1$ is possible
and so there is a ``nonsense fixed-pole'' at $j=1$. The conflict with
unitarity is resolved by the fixed pole mixing with the elementary
gluon and producing Reggeization.

To analyse multiparticle unitarity in the $j$-plane, we need to generalize
all of the elastic scattering formalism. We require

\begin{itemize}

\item Multiparticle, {\it many-variable}, dispersion relations.

The analyticity properties of multiparticle amplitudes are very complicated
but (20 years ago) it was shown\cite{hs,harw,arw1} that, in multi-Regge
asymptotic regions, the necessary dispersion relations hold. This is
sufficient to obtain analytically continued partial-wave amplitudes\cite{arw1}.
Spectral components of the (asymptotic) dispersion relations are labeled by
{\it hexagraphs}. These are tree graphs having the form illustrated in
Fig.~2.1.
\begin{center}
\leavevmode
\epsfxsize=4in
\epsffile{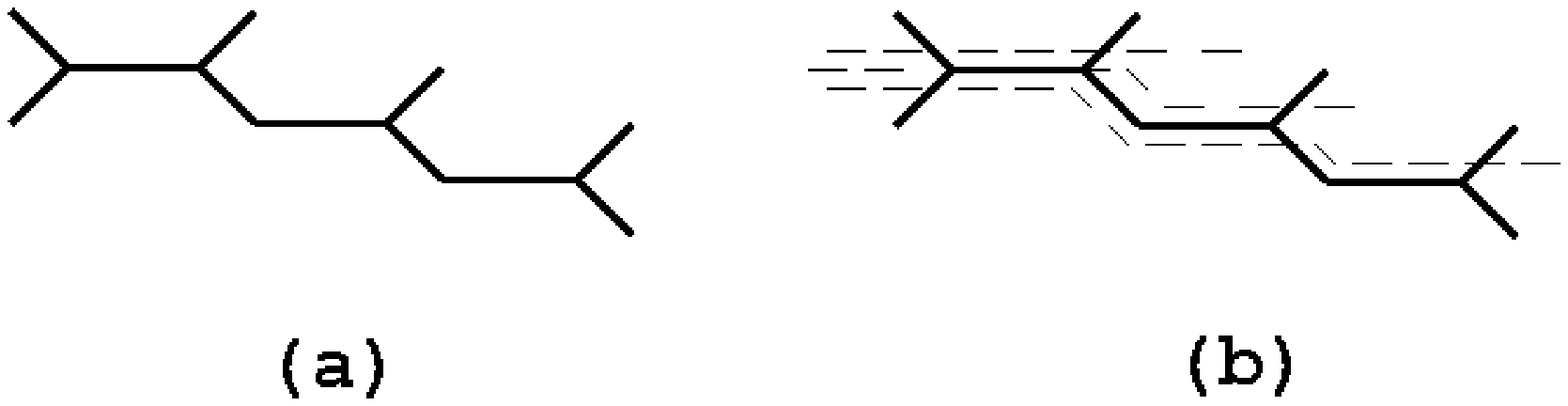}

Fig.~2.1 (a) A hexagraph for the six-particle amplitude (b) cuts through the
hexagraph
\end{center}

Possible cuts through a hexagraph, as illustrated, give the multiple
discontinuities of the spectral component that the graph represents.

\item Continuations to complex {\it angular momenta and helicities.}

For each hexagraph component, distinct continuations are possible and the
hexagraph notation also indicates this. For example, introducing angular
momenta and helicities corresponding to the elements of the hexagraph as in
Fig.~2.2, a continuation can be made to complex $j_1$, $n_2$, and $n_3$
with $j_2-n_2$ and $j_3-n_3$ (which are coupled in the hexagraph) held fixed
at integer values.
The complete set of hexagraph $j$- and $n$-plane continuations are
sufficient\cite{arw1} to write Sommerfeld-Watson transforms and obtain
multi-Regge asymptotic behavior. These continuations are also sufficient to
obtain the $t$-channel unitarity contributions of multi-Regge pole states
that we discuss below. We shall find that {\it reggeon singularities are
generated as Regge cuts} and that {\it particle singularities give reggeon
interactions}.

\begin{center}
\leavevmode
\epsfxsize=2.5in
\epsffile{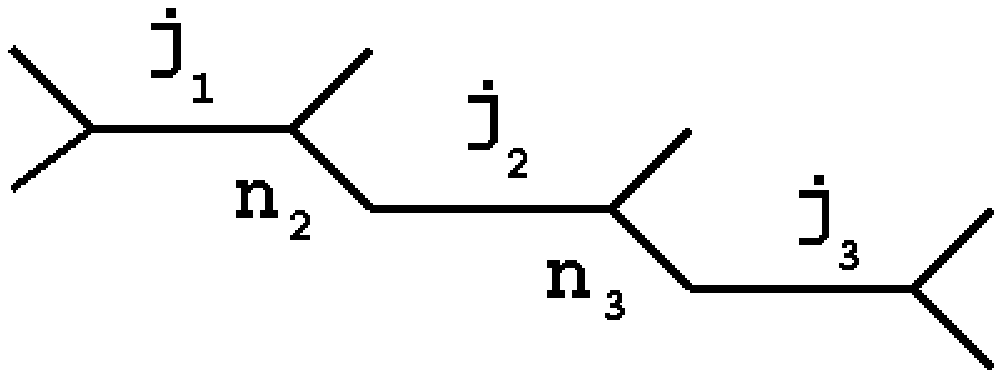}

Fig.~2.2 Angular momenta and helicities associated with a hexagraph.
\end{center}
\end{itemize}

Our ultimate aim is to construct ``Yang-Mills reggeon theories'' by
using $j$-plane unitarity directly. We can by-pass momentum-space
calculations completely by using the following elements.

\begin{description}

\item[{[A]}] Gauge invariance is input via the Ward identity constraint - that
reggeon interactions  vanish at zero transverse momentum.

\item[{[B]}] The ``nonsense'' zero/pole structure required by general
analyticity properties is imposed, in addition to Ward Identity zeroes.

\item[{[C]}] The group structure is input via the triple reggeon vertex.

\item[{[D]}] $t$-channel unitarity is used to determine both $j$-plane
Regge cut discontinuities and particle threshold discontinuities due to
``nonsense'' states.

\item[{[E]}]  The $j$-plane and $t$-plane discontinuity
formulae are expanded simultaneously around $j=1$ and in powers
of $g^2$.

\end{description}

\subhead{2.1 Reggeon Unitarity}

We first go through a 30 year old\cite{gpt,arw1} manipulation of $t$-channel
unitarity which, a-priori, is independent of gauge invariance. Consider the
four-particle intermediate state as illustrated in Fig.~2.3.

\begin{center}
\leavevmode
\epsfxsize=3.5in
\epsffile{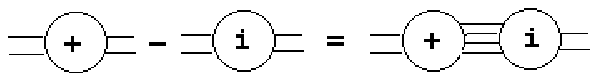}

Fig.~2.3 The four-particle intermediate state.
\end{center}
The $i$ denotes an amplitude evaluated on the unphysical side of the
four-particle branch-cut. (We will avoid discussing subtleties associated
with the definition of $i$ amplitudes, in particular the specification of
the additional boundary-values involved.)  We use multiparticle
partial-wave amplitudes corresponding to the ``coupling scheme''
illustrated in Fig.~2.4.
\begin{center}
\leavevmode
\epsfxsize=2in
\epsffile{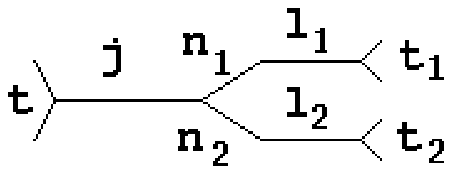}

Fig.~2.4 Partial-wave coupling scheme for the 2-4 production amplitude
\end{center}
$l_1~(l_2)$ and $n_1~(-n_2)$ are respectively the angular momentum
and helicity (in the overall center of mass) of the two-particle state with
invariant energy $t_1~(t_2)$.

The partial-wave projection of Fig.~2.3 is
$$
\eqalign{~~~~~~~~a_j(t)-a^i_j(t)= \int d\rho \sum_{|n_1+n_2|\leq j}
{}~\sum_{l_1\geq
|n_1|}~ \sum_{l_2\geq |n_2|} a_{j\til{l}\til{n}}(t,\til{t})
a^i_{j\til{l}\til{n}}(t,\til{t})}
\auto\label{proj}
$$
where, if all particles have mass $m$ but are not identical,
$$
\eqalign{\int d\rho (t,t_1,t_2)~& =~{i \over (2\pi)^52^6} \int dt_1dt_2\cr
&\times \left[{\lambda^{1/2}(t,t_1,t_2)\over
t}\right]\left[{\lambda^{1/2}(t_1,m^2,m^2)\over t_1}\right]
\left[{\lambda^{1/2}(t_2,m^2,m^2)\over t_2}\right]}
\auto\label{pha}
$$
with the integration region defined by $\lambda~\geq~0$, for each of the
three $\lambda$ functions.

\begin{center}
\leavevmode
\epsfxsize=2.5in
\epsffile{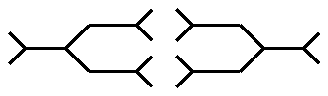}

Fig.~2.5 Hexagraph contributions to the unitarity integral.
\end{center}
Temporarily ignoring signature problems, the
continuation to complex $j$ for the hexagraph contributions of Fig.~2.5
is given by
$$
\eqalign{\sum_{{\scriptstyle  n_1 \geq 0,~ n_2 \geq 0
\atop \scriptstyle  j ~\geq~ n_1+n_2 }}~
{}~\to ~-~ {sin\pi j \over 2^2} \int_{C_j}
{dn_1dn_2 \over sin\pi n_1 sin\pi n_2 sin\pi(j- n_1-n_2)} }
\auto\label{sin}
$$
where the integration contour is defined so that, for $j ~\sim~-1/2$, $C_j
{}~\equiv~
[n_r=-1/4~+i\nu_r~, -\infty < \nu_r < \infty~,~ r=1,2]$.

We consider the contribution of Regge poles as illustrated in Fig.~2.6.

\begin{center}
\leavevmode
\epsfxsize=3.5in
\epsffile{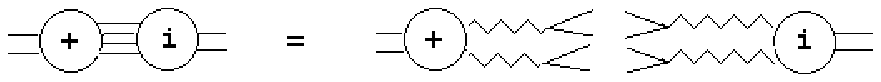}

Fig.~2.6 Regge poles in the production amplitude.
\end{center}
We consider, specifically, $l_1=n_1$ and $l_2=n_2$. Writing
$$
a_{jn_1n_1n_2n_2}(t,t_1,t_2)~=~A_{\alpha_1\alpha_2}~{1 \over [l_1-\alpha_1]
[l_2-\alpha_2]}~\beta_1\beta_2~~~~~\alpha_i=\alpha(t_i)~,
\auto\label{prod}
$$
utilising two-particle unitarity, and picking out the nonsense pole at
$j=n_1+n_2-1$ gives
$$
\eqalign{a_j - a_j^{i} =
{}~-{sin\pi j \over 2^2\pi} ~&\int d\tilde{\rho}
\int_{C_j} {dn_1dn_2 \over sin\pi n_1 sin\pi n_2 (j- n_1-n_2 +1)}\cr
&\times~A_{\til{\alpha}}A^{i}_{\til{\alpha}}
\bigg/ (n_1-\alpha_1)
(n_2-\alpha_2) }
\auto\label{jpl}
$$
where now
$$
\int d\tilde{\rho} ~=~{i \over 2^5\pi^3}
\int  dt_1dt_2\left[{\lambda^{1/2}(t,t_1,t_2)\over
t}\right]
\auto\label{pha1}
$$
Using the threshold behavior
$$
A_{\til{\alpha}}~~~
\centerunder{$\sim$}{\raisebox{-3mm}
{$\scriptstyle \lambda \to 0$}}
{}~~~\biggl[{\lambda(t,t_1,t_2) \over t}\biggr]^{(j-\alpha_1-\alpha_2)}
\auto\label{tild}
$$
we obtain
$$
\eqalign {a_j -a_j^i ~=~ -i~{sin\pi j\over 2^7{\pi}^3}
&\int  {dt_1dt_2 \over \lambda^{1/2}(t,t_1,t_2)} ~A_{\til{\alpha}}
A^i_{\til{\alpha}}\cr
&\times {1 \over sin\pi \alpha_1 sin\pi \alpha_2 (
j-\alpha_1-\alpha_2 +1)} ~~~~+ ...}
\auto\label{rct3}
$$
This leads to the two-reggeon branch-point at $j=2\alpha(t/4)-1 $ generated by
$$
\eqalign{ j=\alpha(t_1)& + \alpha(t_2) -1, ~~~~\lambda(t,t/4,t/4)=0}
\auto\label{2rc}
$$
Since
$$
\int {dt_1dt_2 \over \lambda^{1/2}(t,t_1,t_2)} ~=~2 \int d^2k
\auto\label{jac}
$$
the two-reggeon contribution can naturally be written as a {\it transverse
momentum integral}. The {\it threshold behavior} (\ref{tild})
{\it at the nonsense point} $j=n_1+n_2 -1$ is crucial for this.
Specializing to $j \sim 1$, taking $\alpha(t)= 1 + \Delta(t)
= 1 + \alpha' t + ...$ (and absorbing factors of $\alpha'$ in
$A_{\til{\alpha}}$ and $A^*_{\til{\alpha}}$), gives
for the two-reggeon discontinuity
$$
\delta_{\omega}\biggl\{a_{\omega}\biggr\} ~=~{1 \over 2^3\pi^2}
\int  {d^2k \over k^2 (k-q)^2 } ~A_{\til{\alpha}}
A^*_{\til{\alpha}} \delta(\omega -\Delta_1-\Delta_2)
\auto\label{rct4}
$$
Comparing with (\ref{2rp}) it is clear that introducing a general 2-2 reggeon
interaction will lead to a generalized form of the BFKL equation.

The above analysis of the two-reggeon cut generalises straightforwardly to
the analysis of the N-reggeon cut - which originates from a nonsense state
of N-reggeons i.e. $j= ~\sum_{r=1}^N\alpha_r~ -N +1$. A self-contained set of
reggeon unitarity equations can be written\cite{gpt,arw1} for multireggeon
scattering amplitudes. All the multireggeon discontinuity formulae can be
written in terms of transverse momentum integrals. We emphasize that this is
a property of the phase-space generating the branch-point and is not a
perturbative result.

Until this point we have effectively ignored signature in our discussion of
the two-reggeon cut. However, for the branch-point to actually be generated
there must be no ``nonsense-zero'' of $A_{\til{\alpha}}$ at $j=\alpha_1 +
\alpha_2 - 1$. The dispersion integral representation for partial-wave
amplitudes implies that {\it odd-signature amplitudes have such zeros} and
so the cut appears only in the even signature amplitude.

\subhead{2.2 Reggeization}

Before specializing to a gauge theory we consider, in general, the
``two-reggeon contribution'' in the odd-signature channel that (in the gauge
theory case) will contain the reggeized gluon. We again consider Regge poles
in the four-particle unitarity integral as illustrated in Fig.~2.6. Before
allowing for (square-root) nonsense zeros the $j$-plane contribution is
(with signature effects now included)
$$
{\pi \over 2} sin{\pi \over 2} (j-1) \int d\tilde{\rho}
{A_{\til{\alpha}}A^*_{\til{\alpha}}
\over [j - \alpha_1 -  \alpha_2 +1][sin{\pi\over 2}(\alpha_1-1)]
[sin{\pi \over 2}(\alpha_2-1)]}
\auto\label{rct5}
$$
We focus on the threshold singularity in $t$, which is generated when
$$
\eqalign{\alpha_1~\equiv~&\alpha(t_1)~=~1~,~~~~
\alpha_2~\equiv~\alpha(t_2)~=~1~,\cr
&\lambda(t,t_1,t_2) ~=~0~.}
\auto\label{tth}
$$
We set $\alpha_1=\alpha_2=1$ and consider the leading $t$-dependence for
$j \sim 1$.
Since $j=1$ is the nonsense point relevant for the
phase-space integration, {\it we obtain a transverse momentum integral
- for the leading threshold behavior }. The two-reggeon phase-space gives (for
$\omega = j-1 \sim 0$)
$$
\delta_t\left\{a_{\omega}(t) \right\}
{}~=~{1 \over \omega}~\delta_{q^2}\left\{J_1(q^2)
A_{\til{\alpha}}
A^*_{\til{\alpha}} \right\}
\auto\label{disc2}
$$
where $J_1(q^2)$ is defined by (\ref{j1}). As illustrated in Fig.~2.7,
\begin{center}
\leavevmode
\epsfxsize=4in
\epsffile{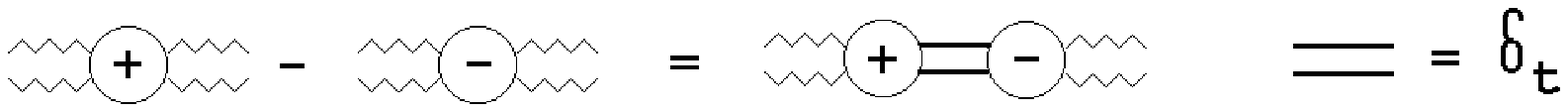}

Fig.~2.7 Discontinuity of the 2-2 reggeon amplitude.
\end{center}
the discontinuity formula obtained holds also for the 2-2 reggeon amplitude
$A_{\til{\alpha}}~\equiv ~~A_{\alpha_1\alpha_2\alpha_3\alpha_4}$
defined at the nonsense point $j=\alpha_1+\alpha_2 -1=\alpha_3+\alpha_4 -1$.

We now consider a gauge theory specifically. We input gauge invariance and the
color structure of the theory, as we have discussed, by requiring

\begin{itemize}

\item Regge pole behavior

\item the color structure of the triple Regge vertex

\item a nonsense zero

\item the Ward identity constraint.

\end{itemize}
The lowest-order form of $A_{\til{\alpha}}$ is then determined to be
$$
A_{\til{\alpha}}~=~{g^2~\sum_{n=1}^N~c_{n~i_1i_2 }c_{n~i_3i_4}~
\omega~q^2 \over (\omega - \Delta(q^2))}
\auto\label{rehc}
$$
The discontinuity formula of Fig.~2.7 gives directly
$$
\Delta(q^2)~=~g^2\sum_{j,k}c^2_{i,j,k}~q^2~ J_1(q^2)
{}~=~g^2N ~q^2~J_1(q^2)
\auto\label{traj1}
$$
which is the familiar leading log form of the gluon trajectory function.

\subhead{2.3 The BFKL Kernel}

We consider the six-particle unitarity integral and analyse it with
partial-wave amplitudes corresponding to the coupling scheme shown in
Fig.~2.8.
\begin{center}
\leavevmode
\epsfxsize=2in
\epsffile{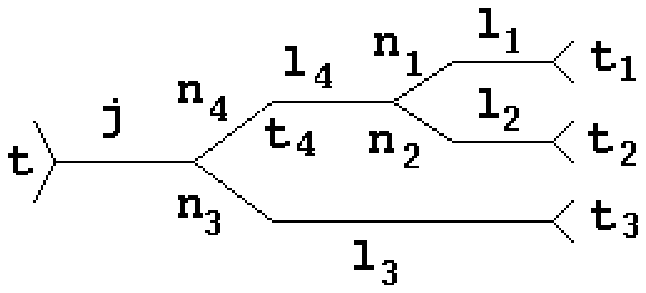}

Fig.~2.8 Coupling scheme for the 2-6 production amplitude
\end{center}
The partial-wave projection of the unitarity integral is
$$
\eqalign{~~~~~~~~a_j(t)-a^i_j(t)= \int d\rho &\sum_{|n_3+n_4|\leq j}
{}~\sum_{|n_1+n_2|\leq l_4}
{}~\sum_{l_1\geq
|n_1|}~ \sum_{l_2\geq |n_2|} ~\sum_{l_3\geq |n_3|}~\sum_{l_4\geq |n_4|}\cr
&\times~
a_{j\til{l}\til{n}}(t,\til{t})
a^i_{j\til{l}\til{n}}(t,\til{t})}
\auto\label{proj6}
$$
The helicity integrals arising from the continuation to complex $j$ of the
helicity sums in (\ref{proj6}) are (from even signature in $j$ and odd
signature in the $n_r$)
$$
\eqalign{ ~~~~~~{1 \over 2^8} sin{ \pi \over 2}j &\int{ dn_3dn_4 \over
sin{\pi \over 2}(j-n_3-n_4)sin{\pi \over 2} (n_3-1)}\cr
&\int { dn_1dn_2 \over
sin{\pi \over 2}(n_4 -n_1 -n_2 +1)sin{ \pi \over 2}(n_1-1)
sin{ \pi \over 2}(n_2-1)}}
\auto\label{6par}
$$
and
$$
\eqalign{\int d\tilde{\rho} (t,t_1,t_2,t_3,t_4)~=~
{\int}_{\lambda(t,t_3,t_4)>0}~ d\tilde{\rho} (t,t_3,t_4)~
{\int}_{\lambda(t_4,t_1,t_2)>0}~ d\tilde{\rho} (t_4,t_1,t_2)}
\auto\label{pha3}
$$

We are interested in the three-particle threshold generated by Regge poles
at $n_i=\alpha_i, ~i=1,2,3$ when
$$
\eqalign{\alpha_1~=~\alpha_2~&=~\alpha_3~=~1~,\cr
\lambda(t_4,t_1,t_2) ~&=~\lambda(t,t_3,t_4) ~=~0~.}
\auto\label{tth1}
$$
and in the two reggeon cut generated by Regge poles at $n_3= \alpha_3$ and
$n_4=\alpha_4$ combining with the nonsense pole at $j=n_3+n_4 -1$. A nonsense
zero prevents a two reggeon cut involving $\alpha_1$ and $\alpha_2$ from
occurring in the $l_4$ channel and so no three reggeon cut is generated in
the overall $j$-plane. Nevertheless, for $j \sim 1$ we have
$$
\alpha_4~\sim~j-\alpha_3 +1 ~\sim~2 - \alpha_3 ~\sim~1~\sim \alpha_1 +
\alpha_2 -1
\auto\label{nons}
$$
and so the nonsense condition $l_4=n_4=n_1+n_2-1$ is satisfied (even though
no two reggeon cut is generated). This second condition
holds in addition to the
$j=n_3+n_4-1$ nonsense condition required for the Regge cut. Since both
conditions hold, threshold factors combine to give the right jacobian
factors to change to transverse momentum variables. ( This implies that
in the following derivation, {\it the BFKL kernel arises entirely from
nonsense states}.) The three-particle discontinuity is then
$$
\delta_t\left\{a_{\omega}(t) \right\}
{}~=~\delta_{q^2}\left\{J_2(q^2)A_{\til{\alpha}}
A^*_{\til{\alpha}} \right\}
\auto\label{ttth}
$$
where
$$
\eqalign{J_2(q^2) ~
=~ {1 \over (16\pi^3)^2} ~\int {d^2k_1d^2k_3 \over k^2_1k^2_3(q-k_1-k_3)^2}}
\auto\label{J2}
$$
There is a factor of $\omega^{-1}$ missing compared to (\ref{disc2}) because
we have extracted nonsense zeroes from the amplitudes.

The lowest-order two-particle/three-reggeon amplitude is determined by
factorization. Since $n_4=(j-\alpha_3+1)$ and
$(n_4-\alpha_4)=(\omega - \Delta_3 - \Delta_4)$ we have

\noindent \parbox{.6in}{
$$
A_{\til{\alpha}}=
$$}
\parbox{3.1in}{
\begin{center}
\leavevmode
\epsfxsize=3in
\epsffile{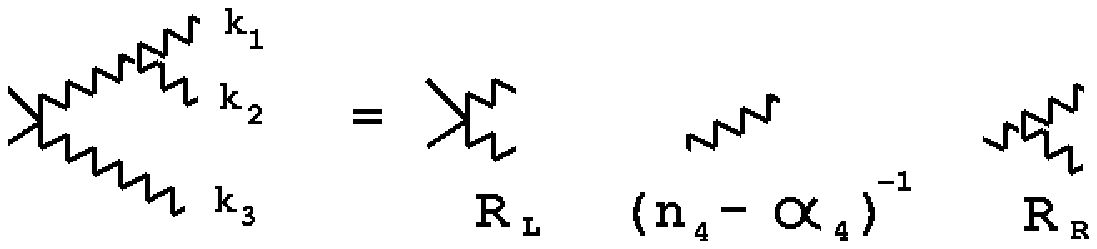}
\end{center}}
\parbox{2.3in}{
$$
{}~\sim ~{ R_L~ gc_{ijk} \over
\omega - \Delta_3 - \Delta_4}
\auto\label{rpf}
$$}
where $R_R$ is the triple reggeon vertex
(except that since we have extracted a nonsense zero there is no momentum
factor) and $R_L$ is an external vertex which we can take to be a constant
carrying zero color i.e. we write $R_L~=~\delta_{ij}$.

Working to $O(g^2)$ in the overall discontinuity and summing over colors
we obtain
$$
\eqalign{ \delta_{q^2}\biggl\{ a_j(q^2)\biggr\} ~&=
{ g^2~N \over (16\pi^3)^2}
\delta_{q^2} \bigg\{ \int {d^2k_1d^2k_3 \over k^2_1k^2_3(q-k_1-k_3)^2}
{1 \over (\omega - \alpha'k_3^2 - \alpha'(q-k_3)^2)^2}
\biggr\}\cr
&={ g^2~N \over (16\pi^3)}
{}~\delta_{q^2} \bigg\{ \int {d^2k_3 \over k^2_3}
{J_1((q-k_3)^2) \over (\omega - \alpha'k_3^2 - \alpha'(q-k_3)^2)^2}
\biggr\} }
\auto\label{dis3r2}
$$
This is the discontinuity of the reggeon diagram shown in Fig.~2.10

\begin{center}
\leavevmode
\epsfxsize=2in
\epsffile{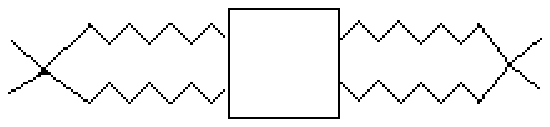}

Fig.~2.10 A reggeon diagram
\end{center}
if the reggeon interaction is the disconnected part of the BFKL kernel.

We must also consider the off-diagonal product of reggeon diagrams shown in
Fig.~2.11.

\begin{center}
\leavevmode
\epsfxsize=1.75in
\epsffile{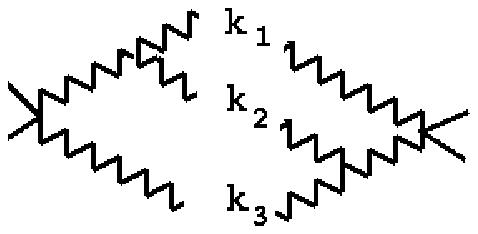}

Fig.~2.11 An off-diagonal product of reggeon diagrams.
\end{center}
The right-hand amplitude has a simple form in the partial-wave coupling
scheme illustrated in Fig.~2.12. Unfortunately, this partial-wave projection
is quite distinct and it is non-trivial to express
the new amplitude in the coupling sceme of Fig.~2.8.

\begin{center}
\leavevmode
\epsfxsize=2.25in
\epsffile{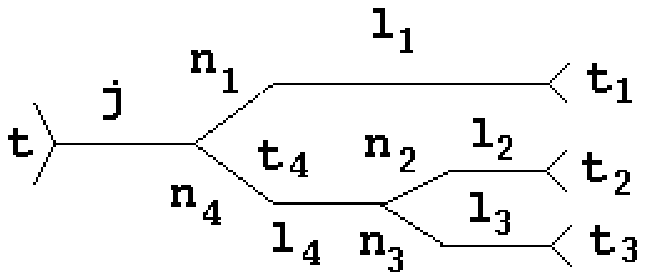}

Fig.~2.12 Alternative coupling scheme
\end{center}

However, if we consider the leading threshold behavior at $t= q^2 = 0$,
there is a simplification. To
obtain $q^2=0$ from three ``massless'' particles, i.e. with $k_i^2 = 0 ,i=
1,2,3,$ {\it all three momenta must be parallel.} This implies that {\it in
this
special case} the relevant variables of Figs.~2.8 and 2.12 degenerate. The
helicities of the three particles can be identified, the angles conjugate to
$j$ and $n_4$ can essentially be identified within each scheme and also in
the two schemes. In this special kinematic configuration we can write

\noindent \parbox{2in}{
\begin{center}
\leavevmode
\epsfxsize=1in
\epsffile{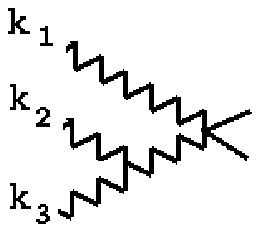}
\end{center}}
\parbox{4in}{
$$
\sim~~~~{R_L~R_R \over \omega - \Delta^*_1 -
\Delta^*_{23}}~~~~~~~~~~~~~~~~~~
\auto\label{rpf2}
$$}
where $\Delta_{23} = \alpha'(k_2 + k_3)^2$, $R_R~=~\delta_{ij}$
and $R_L$ is the triple reggeon vertex.

Combining (\ref{rpf2}) and (\ref{rpf}) and inserting in (\ref{ttth}) we
again obtain a reggeon diagram of the form of Fig.~2.10. Adding the two
possible off-diagonal products we obtain the forward connected BFKL kernel
(\ref{int}). The sign is determined by a detailed discussion of the
helicities of the reggeons involved. The remaining $(k_1 + k_2)^2$
component has no discontinuity in $q^2$ and can not be determined by
unitarity. It is immediately determined as the first correction away from
$q^2=0$ once we impose the Ward identity constraint that is our
input of gauge invariance. Therefore the full, conformally invariant , BFKL
kernel is determined by the combination of $t$-channel unitarity and Ward
identity constraints.

\subhead{2.4 $O(g^4)$ 2-2 Reggeon Interactions}

We study the eight-particle intermediate state and consider the reggeon
contributions shown in Fig.~2.13.

\begin{center}
\leavevmode
\epsfxsize=2.5in
\epsffile{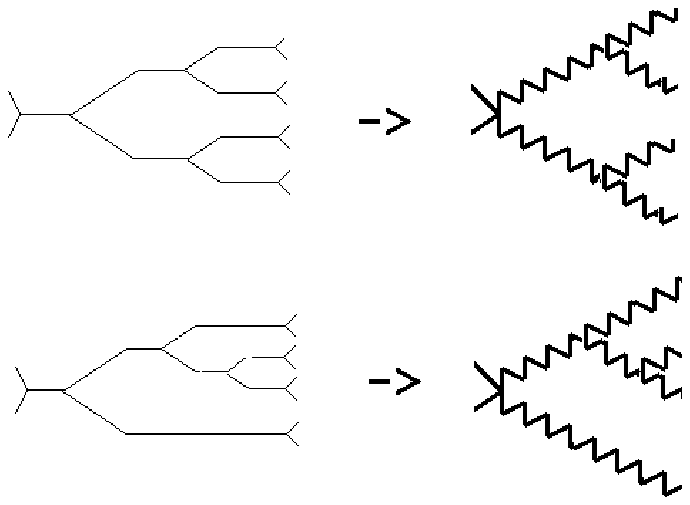}

Fig.~2.13 Reggeon contributions to eight-particle unitarity.
\end{center}
Naively we might expect the previous analysis to generalise
straightforwardly as illustrated in Fig.2.14.

\begin{center}
\leavevmode
\epsfxsize=4in
\epsffile{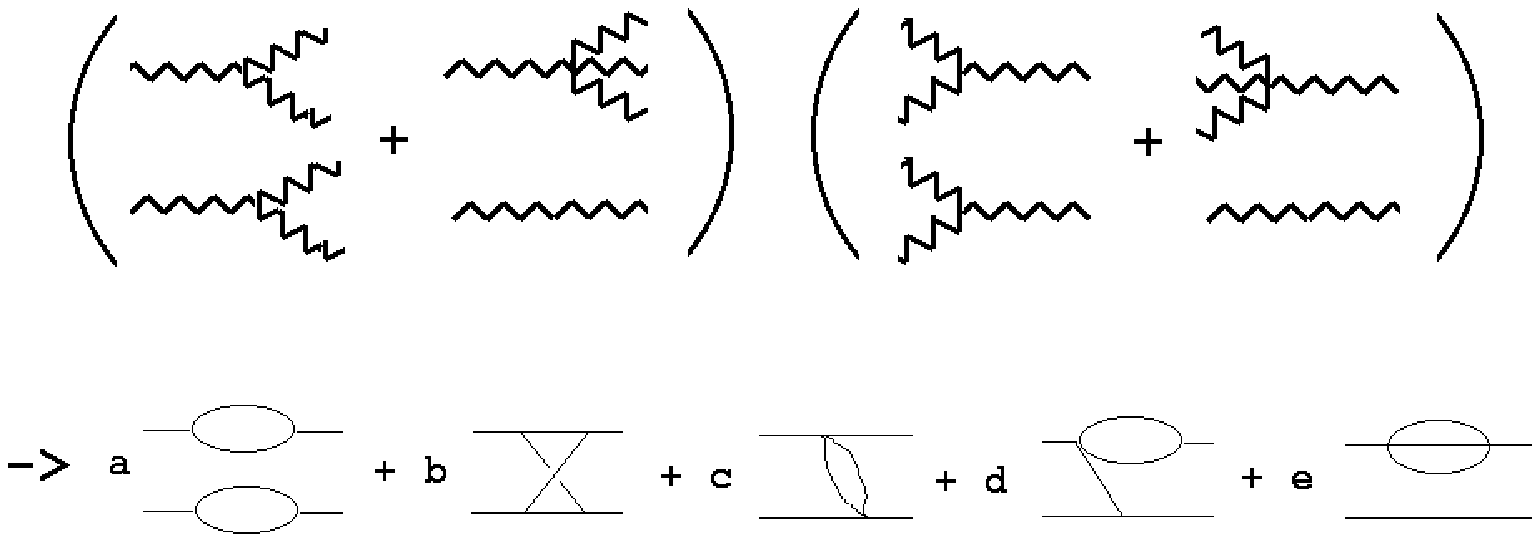}

Fig.~2.14 Reggeon interactions.
\end{center}
This would be the kernel given by the reggeon diagram analysis. The
coefficients $a,b,c,d,e$ are determined by the Ward identity and infra-red
finiteness constraints and might be expected to emerge simply from the
unitarity analysis. It is not so simple. We can summarise the subtleties as
follows.

\begin{description}

\item[i)] The diagram (with coefficient) $a$ is not present, it can be
reduced to a sum of reggeization contributions. In fact this diagram
requires a minimum rapidity cut-off for it's definition.

\item[ii)] $c,d$ and $e$ all involve the 1-3 reggeon coupling (which in
principle could be zero). As a result nonsense conditions do not follow and
only the combination of infra-red finiteness and Ward identity constraints
implies that all the diagrams are present
as transverse momentum integrals in the infra-red region.

\item[iii)] In the infra-red region, diagram {\it $b$ directly generates a
transverse momentum integral} but only {\it for the leading threshold behavior
in the reggeon mass variables.} Also, the product of distinct partial-wave
amplitudes involved generates an {\it overall normalization ambiguity in
transforming from one partial-wave to the other}.

\end{description}

We shall see in the next Section that the component of diagram $b$ that
emerges as most unambiguously defined indeed has special importance.

\mainhead{3. PROPERTIES OF THE $O(g^4)$ KERNELS}

We now return from the unitarity analysis to the kernels that we initially
constructed using reggeon diagrams\cite{ker}. We first discuss the properties
of these kernels and then return to the issue of how
they contribute as higher-order corrections to the BFKL kernel.
As we outlined above, and is described in more detail in
\cite{ker}, the {\it construction procedure} is to use reggeon diagrams to
generate all possible transverse momentum diagrams and then use Ward
identity and infra-red finiteness constraints to determine the relative
coefficients.

\subhead{3.1 The $O(g^4)$ 2-4 Kernel}

We discuss this only briefly. The complete 2-4 kernel is given by
$$
\eqalign{& K^{(4)}_{2,4}(k_1,k_2,k_3,k_4,k_5,k_6)~=~
{}~\sum_{\scriptscriptstyle 1<->2}
2\pi^3k_2^2\Biggl(\delta^2(k_2-k_6)K^{(4)}_{1,3}(k_1,k_3,k_4,k_5)\cr
&~+~\delta^2(k_2-k_5)K^{(4)}_{1,3}(k_1,k_3,k_4,k_6)
{}~+~\delta^2(k_2-k_4)K^{(4)}_{1,3}(k_1,k_3,k_5,k_6)\cr
&~+~\delta^2(k_2-k_3)K^{(4)}_{1,3}(k_1,k_4,k_5,k_6)\Biggr)
-~K^{(4)}_{2,4}(k_1,k_2,k_3,k_4,k_5,k_6)_c}
\auto\label{k424}
$$
The last term $K^{(4)}_{2,4}(k_1,..,k_6)_c$ is the connected part of the
kernel and is generated by the reggeon diagrams shown in Fig.~3.1
\begin{center}
\leavevmode
\epsfxsize=4.5in
\epsffile{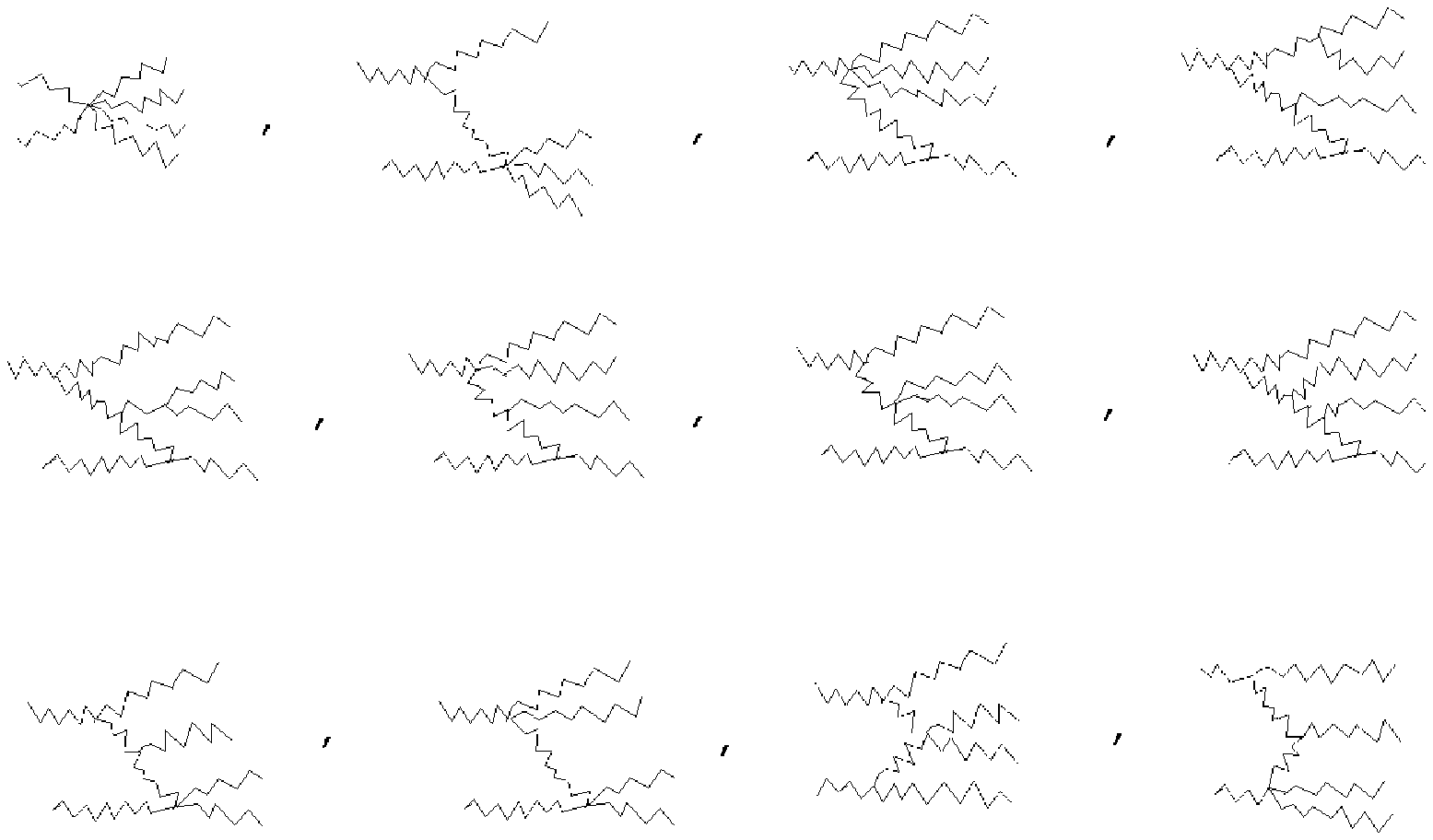}

Fig.~3.1 Reggeon diagrams for the connected 2-4 reggeon kernel
\end{center}
The resulting transverse momentum diagrams have already appeared in Fig.~1.7.
In detail we have
$$
\eqalign {& g^{-4}K^{(4)}_{2,4}(k_1,k_2,k_3,k_4,k_5,k_6)_c~=~
\sum_{\scriptscriptstyle 1<->2}~\Biggl(~(k_1 + k_2)^2
- \Biggl( {k_1^2(k_4+k_5+k_6)^2 \over (k_1-k_3)^2}\cr
&~+~{k_1^2(k_3+k_5+k_6)^2 \over (k_1-k_4)^2}
{}~+~{k_1^2(k_3+k_4+k_6)^2 \over (k_1-k_5)^2}
{}~+~{k_1^2(k_3+k_4+k_5)^2 \over (k_1-k_6)^2}\Biggr)\cr
&~-~{1 \over 4}\Biggl( {k_1^2k_3^2 \over (k_2-k_3)^2}
{}~+~{k_1^2k_4^2 \over (k_2-k_4)^2}
{}~+~{k_1^2k_5^2 \over (k_2-k_5)^2}
{}~+~{k_1^2k_6^2 \over (k_2-k_6)^2}\Biggr)\cr
&~+~{1 \over 2}\Biggl( {k_1^2(k_5+k_6)^2 \over (k_2-k_5-k_6)^2}
{}~+~{k_1^2(k_5+k_4)^2 \over (k_2-k_5-k_4)^2}
{}~+~{k_1^2(k_4+k_6)^2 \over (k_2-k_4-k_6)^2}\cr
&~+~{k_1^2(k_3+k_6)^2 \over (k_2-k_3-k_6)^2}
{}~+~{k_1^2(k_5+k_3)^2 \over (k_2-k_5-k_3)^2}
{}~+~{k_1^2(k_3+k_4)^2 \over (k_2-k_3-k_4)^2}\Biggr)\cr
&~+~{1 \over 2}\Biggl( {k_1^2k_2^2(k_4+k_5)^2 \over (k_1-k_3)^2(k_2-k_6)^2}
{}~+~{k_1^2k_2^2(k_3+k_5)^2 \over (k_1-k_4)^2(k_2-k_6)^2}
{}~+~{k_1^2k_2^2(k_3+k_4)^2 \over (k_1-k_5)^2(k_2-k_6)^2}\cr
&~+~{k_1^2k_2^2(k_3+k_6)^2 \over (k_1-k_4)^2(k_2-k_5)^2}
{}~+~{k_1^2k_2^2(k_4+k_6)^2 \over (k_1-k_3)^2(k_2-k_5)^2}
{}~+~{k_1^2k_2^2(k_5+k_6)^2 \over (k_1-k_3)^2(k_2-k_4)^2}\Biggr)\cr
&~-~{1 \over 4}\Biggl( {k_1^2k_2^2k_4^2 \over (k_1-k_3)^2(k_2-k_5-k_6)^2}
{}~+~{k_1^2k_2^2k_5^2 \over (k_1-k_3)^2(k_2-k_4-k_6)^2}\cr
&~+~{k_1^2k_2^2k_6^2 \over (k_1-k_3)^2(k_2-k_4-k_5)^2}
{}~+~{k_1^2k_2^2k_3^2 \over (k_1-k_4)^2(k_2-k_5-k_6)^2}\cr
&~+~{k_1^2k_2^2k_5^2 \over (k_1-k_4)^2(k_2-k_3-k_6)^2}
{}~+~{k_1^2k_2^2k_6^2 \over (k_1-k_4)^2(k_2-k_3-k_5)^2}\cr
&~+~{k_1^2k_2^2k_4^2 \over (k_1-k_5)^2(k_2-k_3-k_6)^2}
{}~+~{k_1^2k_2^2k_3^2 \over (k_1-k_5)^2(k_2-k_4-k_6)^2}\cr
&~+~{k_1^2k_2^2k_6^2 \over (k_1-k_5)^2(k_2-k_4-k_3)^2}
{}~+~{k_1^2k_2^2k_4^2 \over (k_1-k_6)^2(k_2-k_5-k_3)^2}\cr
&~+~{k_1^2k_2^2k_5^2 \over (k_1-k_6)^2(k_2-k_4-k_3)^2}
{}~+~{k_1^2k_2^2k_3^2 \over (k_1-k_6)^2(k_2-k_4-k_5)^2} \Biggr)\Biggr)}
\auto\label{k424c}
$$
The remaining terms in (\ref{k424}) are disconnected components and involve
the 1-3 reggeon interaction $K^{(4)}_{1,3}(k,k_1,k_2,k_3)$,
which is given by the reggeon diagrams of Fig.~3.2

\begin{center}
\leavevmode
\epsfxsize=3.5 in
\epsffile{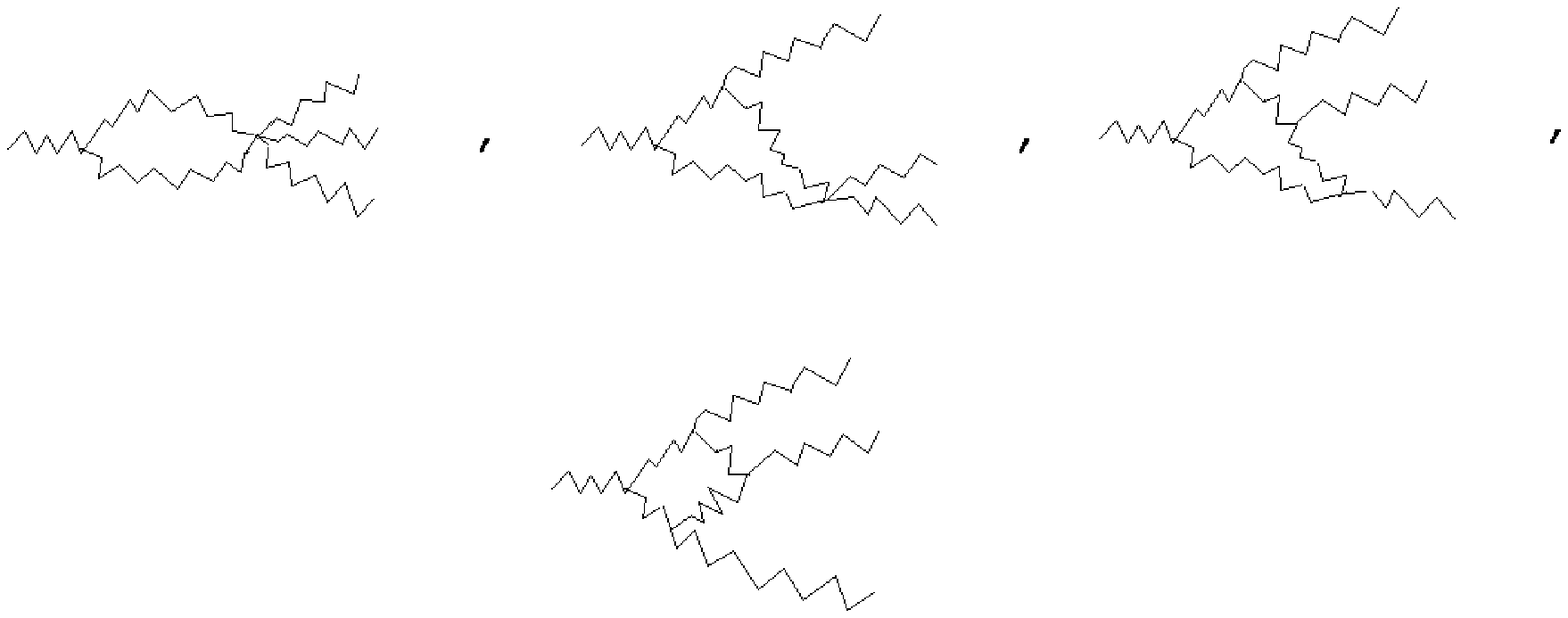}

Fig.~3.2 Reggeon diagrams giving $K^{(4)}_{1,3}$.
\end{center}
and can be written as
$$
\eqalign{ K^{(4)}_{1,3}(k,k_1,k_2,k_3)~=~{1 \over (2\pi)^3}\int
{d^2k_1 \over k_1^2}{d^2k_2 \over k_2^2}&~k^2\delta^2(k-k_1-k_2)\cr
\times~K^{(4)}_{2,3}(k_1,k_2,k_3,k_4,k_5)}\auto
$$
where
$$
\eqalign{ g^{-4}K^{(4)}_{2,3}(k_1,k_2,k_3&,k_4,k_5)~=~\sum_{\scriptscriptstyle
1<->2}~~\Biggl(~(k_1+k_2)^2
-~\Biggl( {k_1^2(k_4+k_5)^2 \over (k_1-k_3)^2}\cr
&~+~{k_1^2(k_3+k_5)^2 \over (k_1-k_4)^2}
{}~+~{k_1^2(k_3+k_4)^2 \over (k_1-k_5)^2} \Biggr)\
{}~+~{1 \over 3}\Biggl( {k_1^2k_5^2 \over (k_2-k_5)^2}\cr
&~+~{k_1^2k_4^2 \over (k_2-k_4)^2}
{}~+~{k_1^2k_3^2 \over (k_2-k_3)^2} \Biggr)
{}~+~{2 \over 3}\Biggl( {k_1^2k_2^2k_4^2 \over (k_1-k_3)^2(k_2-k_5)^2}\cr
&~+~{k_1^2k_2^2k_5^2 \over (k_1-k_4)^2(k_2-k_3)^2}
{}~+~{k_1^2k_2^2k_3^2 \over (k_1-k_5)^2(k_2-k_4)^2} \Biggr)\Biggr)}
\auto\label{k423}
$$

It is straightforward to check\cite{ker} that all the Ward identity
constraints are satisfied by $K^{(4)}_{2,4}$. This vertex is essentially
that calculated directly by Bartels and W\"usthoff\cite{bw}, although to
obtain precisely the same result it is necessary to include the relevant
color factors correctly. Note that the existence of $K^{(4)}_{2,4}$
immediately implies that there is no closed BFKL equation at
$O(\alpha_s^2)$. To obtain such an equation we have to artificially restrict
the discussion to 2-2 reggeon interactions.

\subhead{3.2 The $O(g^4)$ 2-2 Kernel}

As we discussed earlier, reggeon diagrams containing four-reggeon
intermediate states generate the sum of transverse momentum diagrams for the
2-2 kernel shown in Fig.~1.7 and give five kinematically
distinct terms.
$$
\eqalign{{1 \over (g^2N)^2} K^{(4n)}_{2,2}(k_1&,k_2,k_3,k_4)
{}~=~K^{(4)}_0~+~K^{(4)}_1~+~K^{(4)}_2~+~K^{(4)}_3~+K^{(4)}_4~}.
\auto\label{sum}
$$
with
$$
\eqalign{K^{(4)}_0~=~
\sum ~ k_1^4k_2^4J_1(k_1^2)J_1(k_2^2)(16\pi^3)\delta^2(k_2-k_3)~,}
\auto
$$

$$
\eqalign{K^{(4)}_1~=~-{2 \over 3}~
\sum ~ k_1^4J_2(k_1^2)k_2^2(16\pi^3)\delta^2(k_2-k_3)}
\auto
$$

$$
\eqalign{K^{(4)}_2~=~-  \sum
\Biggl({k_1^2J_1(k_1^2)k_2^2k_3^2+
k_1^2k_3^2J_1(k_4^2)k_4^2 \over
(k_1-k_4)^2} \Biggr),}
\auto
$$

$$
\eqalign{K^{(4)}_3~= ~\sum~
k_2^2k_4^2J_1((k_1-k_4)^2)~,}
\auto
$$
and
$$
\eqalign{K^{(4)}_4~=~{1 \over 2}~\sum~
k_1^2k_2^2k_3^2k_4^2~I(k_1,k_2,k_3,k_4), }
\auto
$$
where $J_1(k^2)$ is defined by (\ref{j1}) and
$$
\eqalign{J_2(k^2)~=~{1 \over 16\pi^3}\int d^2q {1 \over
(k-q)^2}J_1(q^2)~~~,}
\auto
$$

and
$$
\eqalign{ I(k_1,k_2,k_3,k_4)~=~{1 \over 16\pi^3}\int d^2p {1 \over
p^2(p+k_1)^2(p+k_1-k_4)^2(p+k_3)^2}.}
\auto\label{box}
$$

We can demonstrate, diagrammatically, that the Ward identity infra-red
finiteness constraints are satisfied as follows. For an external $k_i$-line

\begin{itemize}

\item $k_i \to 0$ gives zero if the line carrying $k_i$ is the single line of a
1-2,  2-1, or 1-1 vertex.

\item In general, $k_i \to 0$ gives the subdiagram obtained by removing the
line carrying $k_i$.

\end{itemize}
Infra-red divergences occur when the
momentum $k_i$ of an internal line vanishes. If we use a mass regulation,
then, as $m^2~\to ~0$, this gives
$$
\int {d^2k_i ~f(k_i) / (k_i^2 + m^2)} ~\to~{1 \over 2} \int {dk_i^2
\over (k_i^2+m^2)}
\int_0^{2\pi} d\theta  ~f(0) ~\to~\pi~logm^2~f(0)
\auto\label{ird}
$$
where (apart from a factor of $(16\pi^3)^{-1}$) $f(0)$ is obtained from the
original diagram by removing the line carrying $k_i$.

The Ward identity constraint is satisfied by the relation

\noindent \parbox{5in}
{\begin{center}
\leavevmode
\epsfxsize=3in
\epsffile{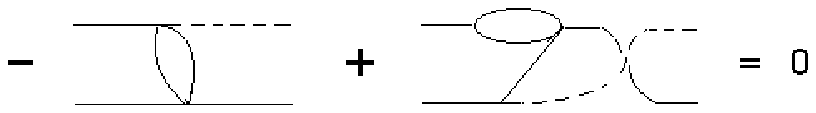}
\end{center}}
\parbox{1in}{
$$
{}~\auto
$$}
(with the notation $-~-~-~\equiv ~k_i \to 0$) and so determines the relative
weight of $K_2$ and $K_3$. There are two infra-red finiteness requirements,
leading to three constraints that determine the relative weights of the
remaining components. First we require that the connected part of the kernel
is infra-red finite before integration. This gives

\noindent \parbox{5in}
{\begin{center}
\leavevmode
\epsfxsize=4in
\epsffile{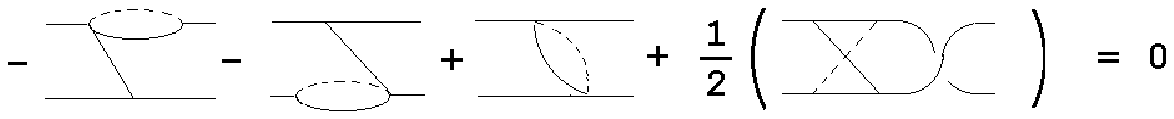}
\end{center}}
\parbox{1in}{
$$
{}~\auto
$$}
and determines $K_4$ relative to $K_2$ and $K_3$. Taking the Ward
identity zeroes into account, infra-red finiteness after integration requires
cancellation, by the disconnected parts, of two divergences due to the
connected part. First the poles of $K_2$ require the cancellation

\noindent \parbox{5in}
{\begin{center}
\leavevmode
\epsfxsize=4in
\epsffile{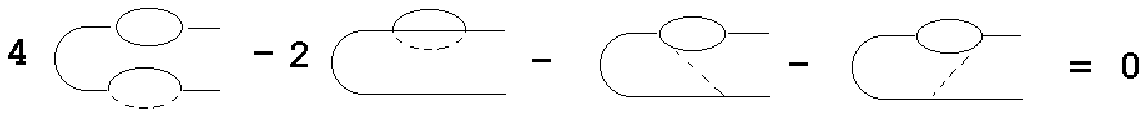}
\end{center}}
\parbox{1in}{
$$
{}~\auto
$$}
Secondly $K_3$ generates a divergence, when both exchanged lines
carry zero transverse momentum, which requires the cancellation

\noindent \parbox{5in}
{\begin{center}
\leavevmode
\epsfxsize=4in
\epsffile{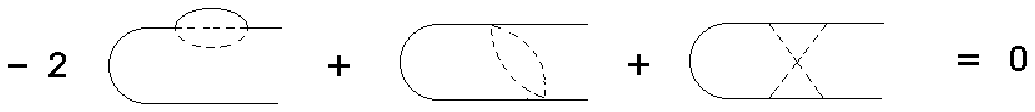}
\end{center}}
\parbox{1in}{
$$
{}~\auto
$$}
This last constraint determines $K^{(4)}_1$ relative to
$K^{(4)}_2 + K^{(4)}_3 + K^{(4)}_4$ and the previous constraint then
determines the relative weight of $K^{(4)}_0$.

The most complicated part of $K^{(4n)}$ is clearly $K^{(4)}_4$ since it
contains the box diagram $I_4$. Using the notation illustrated in Fig.~3.3

\epsffile{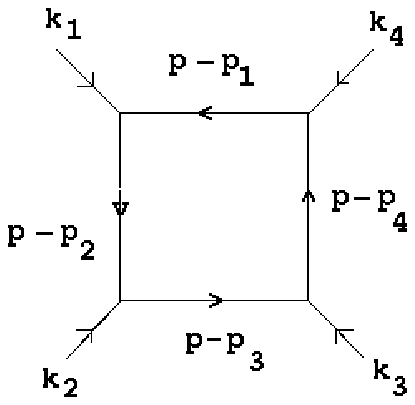}
\begin{center}
Fig.~3.3 Notation for the box diagram
\end{center}
we write
$$
\eqalign{I_4 (p_1&,p_2,p_3,p_4,m^2)~=~\int d^2p ~\Pi_{i=1}^4 ~
{1\over [(p-p_i)^2 - m^2] }}
\auto
$$
$I_4$ can be evaluated\cite{cw1} as a sum of logarithms, i.e.
$$
I_4=\sum_{j\,\,<\,\,k}A_{jk} F_{jk}~
\auto
$$
where the $A_{jk}$ are ``tree-diagrams'' obtained by putting internal lines
$j$ and $k$ on-shell and, writing $p_{jk}=(p_j-p_k)^2$,
$$
F_{jk}~=~{i\pi\over \lambda^{1/2}(p_{jk}^2,m^2,m^2)}Log\Biggl[
{p_{jk}^2 -2 m^2 -\lambda^{1/2}(p_{jk}^2,m^2,m^2)\over
p_{jk}^2 -2 m^2 + \lambda^{1/2}(p_{jk}^2,m^2,m^2)}\Biggr]
\auto
$$
Explicit expressions for the $A_{jk}$ are obtained by introducing dual
vectors to the $p_{jk}$ giving e.g.
\beq
A_{12}={a_{12}\over b_{12}}
\eeq
where
\beqa
a_{12} = &&\biggl[ {{{ k_1} \cdot { k_2}}^2} - { k_1}^{2}\,{ k_2}^{2}
\biggr]\nonumber \\
&&\,\,\,\,\,\,\,\,\,\,\,\times\biggl[{{{ k_1} \cdot { k_2}}^2} -
   { k_1} \cdot { k_2}\,{ k_1} \cdot { k_3} -
   { k_1}^{2}\,{ k_2}^{2} + { k_1}^{2}\,{ k_2} \cdot { k_3}\biggr.
\nonumber \\
&&\biggl. +  ( { k_1} \cdot { k_2} + { k_2}^{2} ) \,
    ( { k_1} \cdot { k_2} - { k_1} \cdot { k_3} + { k_2}^{2} -
      2\,{ k_2} \cdot { k_3} + { k_3}^{2} )\biggr]
\eeqa
\beqa
b_{12} = &&\biggl[ -{{{ k_1} \cdot { k_2}}^2} + { k_1}^{2}\,{ k_2}^{2} +
   {{( { k_1} \cdot { k_2} + { k_2}^{2} ) }^2}\biggr]
\nonumber \\
&&\times\left.\biggl[ -{{( {{{ k_1} \cdot { k_2}}^2} -
         { k_1} \cdot { k_2}\,{ k_1} \cdot { k_3} -
         { k_1}^{2}\,{ k_2}^{2} +
         { k_1}^{2}\,{ k_2} \cdot { k_3} ) }^2}\right.\nonumber \\
&&\left. +
   ( {{{ k_1} \cdot { k_2}}^2} - { k_1}^{2}\,{ k_2}^{2})
      ) \,{{( { k_1} \cdot { k_2} - { k_1} \cdot { k_3} +
         { k_2}^{2} - 2\,{ k_2} \cdot { k_3} + { k_3}^{2} ) }^2}
\right.\biggr]\nonumber \\
\eeqa

In this way we obtain the box diagram as a sum of six logarithms
of two types:

\begin{itemize}

\item[{(1)}] external line ``reggeon mass'' thresholds, $\to$ four logarithms.

\item[{(2)}]``s'' and ``t'' thresholds, $\to$ two
logarithms.

\end{itemize}
The complete kernel can then be written in terms of logarithms with
rational polynomial factors. (In fact a greatly simplified expression can be
found in \cite{cpw}).

\subhead{3.3 The $O(g^4)$ Parton Kernel}

For parton evolution, we require only the much simpler ``forward'' kernel

\noindent \parbox{2in}{
\begin{center}
\leavevmode
\epsfxsize=1.5in
\epsffile{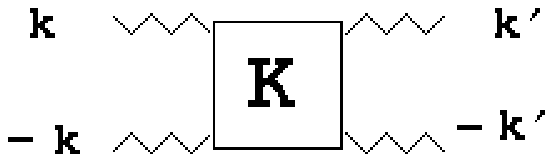}
\end{center}}
\parbox{4in}{
$$
=~K^{(4n)}_{2,2}(k,-k,k',-k')~\equiv ~K^{(4)}(k,k')
\auto
$$}
In the forward direction it is straightforward to combine the type
(2) logarithms from the box with the logarithms
of the connected components $K_2^{(4)}$ and $K_3^{(4)}$, giving
$$
\eqalign{K^{(4)}(k,k')_c~~\centerunder{$\longrightarrow$}{\raisebox{-3mm}
{$\scriptstyle m^2 \to 0$}}
{}~~&{1 \over 8\pi^2}
\Biggl( {k^2{k'}^2 \over (k-k')^2}Log\left[{(k-k')^4 \over k^2{k'}^2
}\right]\cr
&~+~ {k^2{k'}^2 \over (k+k')^2} Log\left[{(k+k')^4 \over k^2{k'}^2 }\right]
\Biggr)~-~\Biggl( ~~{\cal K}_2~~\Biggr)}
\auto\label{4nc}
$$
where
$$
{\cal K}_2~~=~{1 \over
4\pi^2 } ~ {k^2 {k'}^2 (k^2-{k'}^2)\over
(k+k')^2 (k-k')^2}~Log\Biggl[{k^2 \over {k'}^2} \Biggr]
\auto
$$
is {\it separately infra-red finite} as $m^2 \to 0$ and contains
only the type (1) logarithms, i.e. the external reggeon mass thresholds.
${\cal K}_2$ will be very important in the following. {\it It  is
the part of the box-diagram that emerged as a well-defined transverse
momentum integral contribution via the unitarity analysis.}

To obtain the full set of eigenvalues of $K^{(4)}$ we first show
diagrammatically that
$$
K^{(4)} ~=~{1 \over 4} (K_{BFKL})^2 - {\cal K}_2
\auto\label{fbf}
$$
where $K_{BFKL} ~=~ K_{BFKL}(k,-k,k',-k')$ is given by (\ref{bfkl}).
Using the notation of \cite{cw} that $\sum$ indicates summation over all
permutations of initial, final, and intermediate states, we have

\noindent \parbox{1.5in}{
$$
(K_{BFKL})^2~~=~~
$$}
\parbox{3.5in}{
\begin{center}
\leavevmode
\epsfxsize=3in
\epsffile{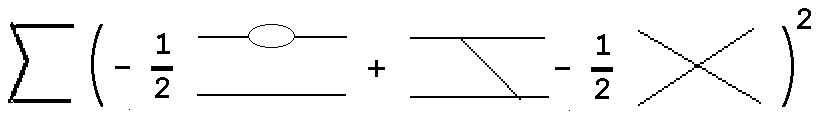}
\end{center}}
\parbox{1in}{
$$
{}~\auto
$$}

\noindent \parbox{5in}{
\begin{center}
\leavevmode
\epsfxsize=4.5in
\epsffile{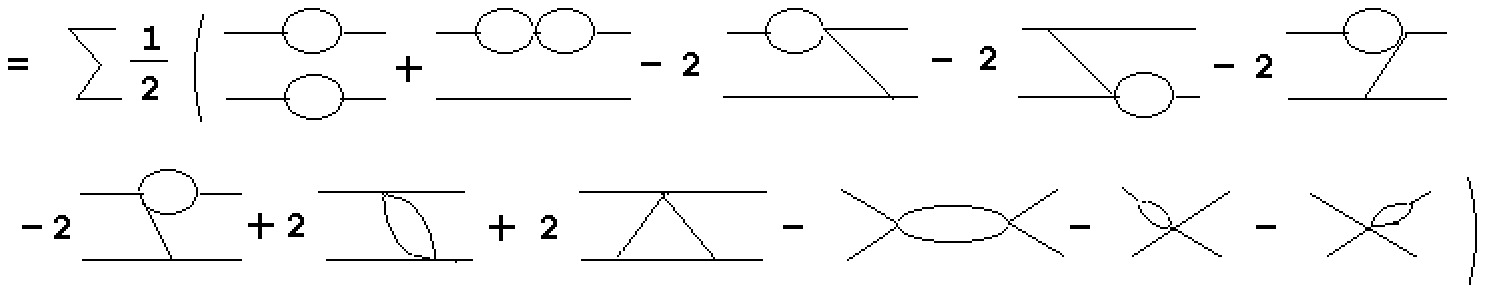}
\end{center}}
\parbox{1in}{
$$
{}~\auto
$$}
Using the forward identities

\noindent \parbox{5in}{
\begin{center}
\leavevmode
\epsfxsize=4.5in
\epsffile{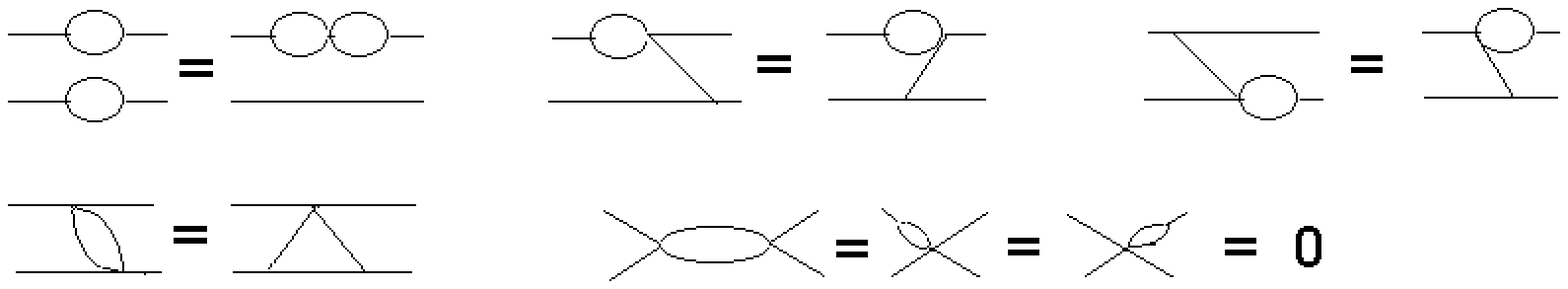}
\end{center}}
\parbox{1in}{
$$
{}~\auto
$$}
then gives

\noindent \parbox{1.5in}{
$$
(K_{BFKL})^2~~=~~
$$}
\parbox{3.5in}{
\begin{center}
\leavevmode
\epsfxsize=3in
\epsffile{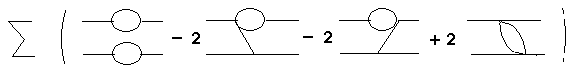}
\end{center}}
\parbox{1in}{
$$
{}~\auto\label{bf2}
$$}
(\ref{fbf}) then follows for the connected components involved. To include
the disconnected parts it is necessary\cite{cw} to utilise the relation
$$
4k^2[J_2(k^2)]~=~3 [k^2J_1(k^2)]^2
\auto
$$
which holds when dimensional regularization is employed.

\subhead{3.4 Eigenvalues and Holomorphic Factorization}

We use the complete set of orthogonal eigenfunctions
$$
\eqalign{ \phi_{\nu,n}(k)~=~({k}^2)^{1/2 +i\nu}~e^{i n \theta}}~~~~~~~~
\nu ~\epsilon~(-\infty, \infty),~~n=0,\pm 1, \pm2, ...
\auto\label{eif}
$$
where $k=(|k|cos\theta,|k|sin\theta)$. The eigenvalues of $K_{BFKL}$ are
${Ng^2 \over 2\pi^2} ~\chi(\nu,n)$ where
$$
\chi(\nu,n)~= ~\psi (1) - Re \psi ({|n|+1 \over 2} +i\nu)
{}~~~~~~~
\auto\label{ei2}
$$
with $\psi (x)~=~{d \over dx}\Gamma(x)$. From (\ref{fbf}) the eigenvalue
spectrum of $K^{(4)}$ is given by $N^2g^4{\cal E}(\nu,n)$ where
$$
{\cal E}(\nu,n)~=~{1 \over \pi} [\chi(\nu,n)]^2~-~\Lambda(\nu,n) ~.
\auto
$$
and $\Lambda(\nu,n)$ are the eigenvalues of ${\cal K}_2$.

To find the $\Lambda(\nu,n)$ we use the dimensionally regularized form of
${\cal K}_2$, i.e.
\beqa
{\cal K}^D_2(k,k')~=~{1 \over 2\pi^2(D-2)} ~ {k^2 {k'}^2 (k^2-{k'}^2)\over
(k+k')^2 (k-k')^2}\left( (k^2)^{D/2~-1} -
({k'}^2)^{D/2~-1}\right). \nonumber \\
\eeqa
We first evaluate
\beqa
&&I_{\theta}[n]\equiv \int_{0}^{2 \pi} d\theta'{ e^{i n \theta'}
\over 1- z(k,k')sin^2\,\,(\theta- \theta')}\nonumber \\
&&z[{k,k'}]=-{4 k^2{k'}^2\over (k^2 -{k'}^2)^2}
\eeqa
where $cos\theta = k\cdot \hat{x}$ and $cos\theta' ={k'}\cdot \hat{x}$.
We get by residui (for $n>\,\,-1$)
\beqa
&& I_{\theta}[n]=-4 i e^{i n \theta} \oint d w {w^{n+1}\over z w^4
+2 (2 - z)w^2 +z} \nonumber \\
&&=\,\,\,\,\,\,2 \pi\delta_{n,2 M} e^{i n \theta}
\left({k^2-{k'}^2\over k^2 + {k'}^2}\right)
\left[ \left({k\over k'}\right)^n \Theta[k'-k] -
\left( {k'\over k}\right)^n\Theta[k-k']\right]. \nonumber \\
\label{Ith}
\eeqa
$2 M$ is an even integer - this will be important in the following.

It is then straightforward to show that
$$
\int {d^Dk' \over (k')^2} {\cal K}^D_2(k,k')
\phi_{\nu,n}(k')~=~\Lambda(\nu,n)  \phi_{\nu,n}(k)
\auto
$$
where, as $D \to 2$,
$$
\eqalign{ \Lambda(\nu,n)~\to~&{1 \over 2\pi^2(D-2)}
 \biggl(\beta\bigl(|n|/2 +D/2 +\nu - 1/2\bigr)\cr
&~-~\beta\bigl(|n|/2-D/2-\nu + 3/2\bigr)
-~\bigl(\beta\bigl(|n|/2 +D +\nu - 3/2\bigr)\cr
&~+~ \beta\bigl(|n|/2-D -\nu + 5/2 \bigr) \bigr)
\biggr)~,}
\auto
$$
$\beta(x)$ is the incomplete beta function, i.e.
$$
\eqalign{
\beta(x)~&=~\int^1_0 dy~y^{x -1}[1+y]^{-1} \cr
&=~{1\over 2}\biggl(\psi\bigl({x+1\over 2}\bigr) -
\psi\bigl({x\over 2}\bigr)\biggr), }
\auto
$$
and so, at $D=2$,
$$
\Lambda(\nu,n) ~=~-~{1 \over 4\pi}
\biggl(\beta'\bigl({|n| + 1\over 2} +
i\nu\bigr)
{}~+~\beta'\bigl({|n| + 1 \over 2} -i\nu\bigr)\biggr)
\auto\label{lam}
$$
where we can write
$$
\eqalign{\beta'(x)~=~{1\over 4}\biggl(\psi'\bigl({x+1\over 2}\bigr) -
\psi'\bigl({x\over 2}\bigr)\biggr) ~. }
\auto\label{ps1}
$$
with
$$
\eqalign { \psi'(z)~=~\sum_{r=0}^{\infty} {1
\over (r+z)^2}, }
\auto\label{ps2}
$$

Using (\ref{ps1}) we can show that the $\Lambda(\nu,n)$ have the important
property of {\it holomorphic factorization} that is very closely related to
conformal symmetry\cite{kir1}. That is we can write
$$
\Lambda(\nu,n)~=~{\cal G}\bigl[m(1-m)\bigr]~+
{}~{\cal G}\bigl[\tilde{m}(1-\tilde{m})\bigr]
\auto\label{hf}
$$
where $m=1/2 + i\nu + n/2$ and $\tilde{m}= 1/2 + i\nu -n/2$ are conformal
weights. We use
$$
\eqalign{ 16\pi\Lambda(\nu,n) ~=&~-~4\biggl(\beta'\bigl(m \bigr)
{}~+~\beta'\bigl(1-\tilde{m} \bigr)\biggr) \cr
=& ~\psi'\biggl({m+1 \over 2}\biggr) ~-~\psi'\biggl({m \over 2}\biggr)
{}~+~\psi'\biggl({2- \tilde{m} \over 2}\biggr)
{}~-~\psi'\biggl({1- \tilde{m} \over 2}\biggr) \cr
=& ~\sum_{r=0}^{\infty} {1 \over (r + {3 \over 4} + {n \over 4} +
{i\nu \over 2})^2}~-
{}~\sum_{r=0}^{\infty} {1 \over (r + {1 \over 4} + {n \over 4} +
{i\nu \over 2})^2}\cr
+& ~\sum_{r=0}^{\infty} {1 \over (r + {3 \over 4} + {n \over 4} -
{i\nu \over 2})^2}
{}~-~ ~\sum_{r=0}^{\infty} {1 \over (r + {1 \over 4} + {n \over 4} -
{i\nu \over 2})^2} }
\auto\label{lam1}
$$
We next show that this expression is unchanged if we simultaneously send
$m \to 1-m$ and $ \tilde{m} \to 1-\tilde{m}$, i.e. $n \to -n, ~\nu \to -\nu$.
At this point {\it it is crucial that $n$ is an even integer.} Writing $n=2M$,
we obtain
$$
\eqalign{ 16\pi\Lambda(-\nu,-n) ~&=
\sum_{r=0}^{\infty} {1 \over (r + {1 \over 4} + {-M +1 \over 2} -
{i\nu \over 2})^2}~-
{}~\sum_{r=0}^{\infty} {1 \over (r + {1 \over 4} + {-M \over 2} -
{i\nu \over 2})^2}\cr
&+ ~\sum_{r=0}^{\infty} {1 \over (r + {1 \over 4} + {-M +1 \over 2} +
{i\nu \over 2})^2}
{}~-~ ~\sum_{r=0}^{\infty} {1 \over (r + {1 \over 4} + {-M \over 2} +
{i\nu \over 2})^2} }
\auto\label{lam2}
$$
and so
$$
\eqalign{ &16\pi\bigl(\Lambda(-\nu,-n) ~-~\Lambda(\nu,n)\bigr)\cr
{}~~~&=~
\sum_{s=-M}^{-1} {1 \over (s + {1 \over 4} + {M +1 \over 2} -
{i\nu \over 2})^2}~-
{}~\sum_{s=-M}^{-1} {1 \over (s + {1 \over 4} + {M \over 2} -
{i\nu \over 2})^2}\cr
&+ ~\sum_{s=-M}^{-1} {1 \over (s + {1 \over 4} + {M +1 \over 2} +
{i\nu \over 2})^2}
{}~-~ ~\sum_{s=-M}^{-1} {1 \over (s + {1 \over 4} + {M \over 2} +
{i\nu \over 2})^2} \cr
&=~\sum_{t=-M/2}^{M/2~-1} {1 \over (t + {3 \over 4} - {i\nu \over 2})^2}~-
{}~\sum_{t=-M/2}^{M/2~-1} {1 \over (-t  - {1 \over 4} - {i\nu \over 2})^2}\cr
&+ ~\sum_{t=-M/2}^{M/2~-1} {1 \over (t + {3 \over 4} + {i\nu \over 2})^2}
{}~-~ ~\sum_{t=-M/2}^{M/2~-1} {1 \over (-t - {1\over 4} + {i\nu \over 2})^2}
\cr
&=~~ 0 }
\auto\label{lam3}
$$
{}From this symmetry, we can write
$$
\eqalign{ 16\pi\Lambda(\nu,n) ~&=~-~2\biggl(\beta'\bigl(m \bigr) ~+~
\beta'\bigl(1-m \bigr)~+~\beta'\bigl(1-\tilde{m} \bigr)\biggr)
{}~+~\beta'\bigl(\tilde{m} \bigr)\biggr)\cr
&\equiv~~ {\cal G}\bigl[m(1-m)\bigr]~+
{}~{\cal G}\bigl[\tilde{m}(1-\tilde{m})\bigr]}
\auto\label{hf1}
$$
as required.

We conclude that ${\cal K}_2$ shares many of the nice properties of the
leading-order $BFKL$ kernel. It is infra-red finite, scale-invariant and has
a {\it new eigenvalue spectrum} satisfying {\it holomorphic factorization}.
It is very interesting to ask whether there is a new conformally invariant,
non-forward, kernel associated with ${\cal K}_2$. (In fact it is shown in
\cite{cw} that ${\cal K}_2$ is the forward component of a new partial-wave
amplitude that appears for the first time at $O(g^4)$ and in \cite{cpw} a
candidate for the non-forward conformally invariant kernel is constructed).

\subhead{3.5 Numerical Evaluation}

We consider now the numerical significance of the eigenvalues of $K^{(4)}$.
The leading eigenvalue is at $\nu=n=0$, as it is for the $O(g^2)$ kernel.
Using the reggeon diagram normalization,
the correction to $\alpha_0$ is given\cite{cw1},
by
$$
9g^4{\cal E}(0,0)/(16\pi^3)~.
\auto
$$
Since
$$
\eqalign{\Lambda(0,0) ~=& ~-~{1 \over 2\pi}\beta'(1/2)\cr
=& ~-~{1 \over 8\pi}\biggl(~\sum_{r=0}^{\infty} {1
\over (r+~1/4)^2} ~- ~\sum_{r=0}^{\infty} {1 \over (r+~3/4)^2}\biggr)\cr
&=~-~{1 \over 8\pi}\biggl(~16~+~{16 \over 25}~+~{16 \over 81}
{}~+~...~-~{16 \over 9}~-~{16 \over 49}~+~...\biggr)\cr
{}~&\sim ~-~ {1.81~ \over \pi}}
\auto
$$
we obtain from ${\cal K}_2$ alone
$$
\eqalign{ {9g^4\over 16\pi^3}\Lambda(0,0)
{}~\sim~-16.3 {{\alpha_s}^2 \over \pi^2} }
\auto
$$
The complete $K^{(4n)}$ gives
$$
\eqalign{
{}~\sim&~{9g^4 \over 16 {\pi}^4}\biggl([2ln2]^2
{}~-~1.81 \biggr)\cr
{}~\sim&~{9g^4 \over 16{\pi}^4}\times 0.11 ~\sim~ {{\alpha_s}^2 \over
\pi^2} }
\auto
$$
giving a very small positive effect.

At this point we note that the disconnected part of $K^{(4)}$
contains diagrams, the first kind appearing in (\ref{bf2}), which
can not be interpreted in terms of reggeization
effects. Since reggeization is the only consistent interpretation of
disconnected pieces, these diagrams can not be present in the full
kernel. Elimination of the unwanted diagrams, while
retaining scale-invariance, gives\cite{cw1} uniquely
$$
\eqalign{ \tilde{K}^{(4)} ~=~K^{(4n)}~-~\biggl(K_{BFKL}\biggr)^2~, }
\auto\label{k_4s}
$$
This is a consistent scale-invariant $O(g^4)$ kernel which can be added to
the $O(g^2)$ kernel. In this case, we replace
${\cal E}(\nu,n)$ by $\tilde{{\cal E}}(\nu,n)$ where
$$
\tilde{{\cal E}}(\nu,n)~=~-~{3 \over \pi} [\chi(\nu,n)]^2~-~\Lambda(\nu,n) ~.
\auto
$$
This gives, as a modification of $\alpha_0$,
$$
\eqalign{9g^4{\tilde{{\cal E}}(0,0) \over 16\pi^3}
\sim &~{9g^4 \over 16{\pi}^4}\times  (-5.76  - 1.81)\cr
\sim&~-68 {{\alpha_s}^2 \over \pi^2}}
\auto
$$
which is a substantial negative correction - of the order of $50\%$.

Unfortunately as we have discussed in the last Section there is, even in the
best determined component ${\cal K}_2$, an overall normalization uncertainty
which reduces the immediate significance of these numerical estimates.

\mainhead{4. THE $O(g^4)$ KERNEL FROM THE $s$-CHANNEL EFFECTIVE ACTION}

Kirschner\cite{kir} has discussed the relationship of the ``$t$-channel''
reggeon diagram construction of non-leading kernels to the ``$s$-channel''
multi-Regge effective action\cite{kls} derived from the leading-log
approximation. The full effective action is written as a sum of components
$$
{\cal L}~=~{\cal L}_{kin}~+~{\cal L}_s~+~{\cal L}_p~+~{\cal L}_t
\auto
$$
${\cal L}_t$ contains the triple-gluon vertex for longitudinal gluon fields
$A_+,~A_-,$ describing ``$t$-channel'' exchanged gluons (``$s$-channel''
produced gluons are described by $A_{\perp}$ fields), i.e.
$$
{\cal L}_t ~=~{ig \over 2} \partial \partial^* A^a_-(\partial^{-1}_+A_+T^a
A_+) ~+~\partial \partial^* A^a_+(\partial^{-1}_-A_-T^a
A_-)
\auto
$$
In momentum space the triple vertex has the form
$$
igc_{abc}{(k_1 +k_2)^2 \over k_{2+}} ~~+~~(k_2,b)~<->~(k_1,c)
\auto
$$
and is, essentially, the three-reggeon vertex that we use to construct
reggeon diagrams.

Kirschner has shown that graphs involving triple-gluon vertices can be
regarded as reggeon diagrams, {\it if contributions with $s$-channel gluons
close to mass-shell are added}. The $O(g^4)$ kernel we have discussed arises
from the product of interactions shown in Fig.~4.1

\begin{center}
\leavevmode
\epsfxsize=3.5in
\epsffile{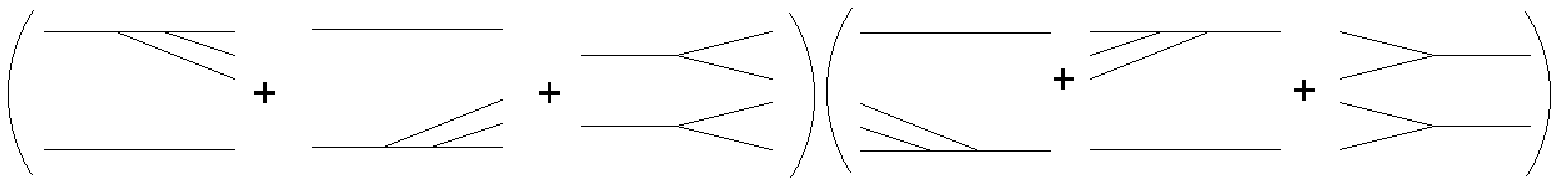}

Fig.~4.1 Reggeon Interactions from the Multi-Regge Effective Action
\end{center}
together with additional contributions from $s$-channel gluons. (This is
clearly analagous to the product of reggeon diagrams illustrated in
Fig.~2.14.) If the
resulting diagrams are written as transverse momentum integrals, the
formalism suggests that the original presence of additional rapidity
integrations produces both

\begin{itemize}

\item {an overall normalization uncertainty}

\item {additional (perhaps slowly varying) transverse momentum dependence.}

\end{itemize}
Since these results are completely consistent with our results, the
effective lagrangian gives a valuable understanding of the reggeon diagram
approximation.

Kirschner also gives an interesting representation for ${\cal K}_2$.
Introducing complex momenta $\kappa$ whose real and imaginary parts are the
two components of conventional transverse momenta
$$
{\cal K}_2(\kappa,\kappa')~=~{2{\kappa}^2{{\kappa}'}^2 \over (2\pi)^3}
\int d^2 {\kappa}'' {1 \over {\kappa}''({\kappa}'' - {\kappa} + {\kappa}')
({\kappa}'' - {\kappa})^*({\kappa}'' + {\kappa}')^*}
{}~~~~+~~c.c.
\auto
$$
This formalism is used in \cite{cpw} to construct the non-forward extension
of ${\cal K}_2$ and is anticipated to be very useful for studying conformal
symmetry properties.

\mainhead{5. CONCLUSIONS}

Used directly, the scale invariant $O(g^4)$ transverse momentum kernel gives
a {\it large reduction of the BFKL small-x behavior of parton
distributions}. However, both $t$-channel unitarity and the multi-Regge
effective action imply that the introduction of {\it scales will modify the
normalization} and significantly modify the kernel at large
$q^2,~k^2,~{k'}^2$. Indeed the outcome of the non-leading $t$-channel unitarity
that we have outlined in Section 2 can be compactly summarized\cite{cw} by
writing, for the full kernel $K_{2,2}(q,k,k')$,
$$
K_{2,2}(q,k,k')~\centerunder{$\longrightarrow$}{\raisebox{-3mm}
{$\scriptstyle q^2, k^2, k'^2 \to 0 $}}~~ g^2 K_{BFKL} +
O(g^4) (K_{BFKL})^2 + O(g^4) {\cal K}_2 ~
\auto\label{suma}
$$
indicating that both the overall normalization and the relative normalization
of the new ${\cal K}_2$ kernel to $(K_{BFKL})^2$, are not determined.

A reggeon interaction derived from $t$-channel unitarity, is necessarily
scale-invariant and only an infra-red approximation. Extrapolation away from
the infra-red region is controlled by the Ward identity constraints and in
\cite{cw1} we conjecture that these constraints lead to conformal
invariance. The BFKL kernel, the triple Regge kernel\cite{bw,ker,blw}, and
the ${\cal K}_2$ kernel we have derived, are the only interactions
studied so far and existing results are consistent with this conjecture.

In \cite{uni} we have outlined a program whereby the scale-dependence of
non-leading reggeon amplitudes can be studied via the Ward identity
constraints. We hope to study this possibility in the future. Of course,
completion of the full $O(\alpha_s^2)$ calculation\cite{fl} should greatly
clarify the role of scale dependence in the 2-2 kernel. Comparison
with the reggeon diagram formalism may then suggest how yet higher-order
contributions can be suitably approximated.

\vspace{.7in}

\centerline{\bf Acknowledgements}

We are grateful to J. Bartels, R. Kirschner and L. Lipatov for valuable
discussions of this work.

\end{document}